\documentclass[twocolumn]{aastex631}
\usepackage{graphicx}	
\usepackage{amsmath}	
\usepackage{amssymb}

\newcommand{\co}{$^{12}$CO~}

\def \coo {C$^{18}$O~}
\def \nthp {N$_{2}$H$^{+}$~}

\newcommand{\kms}{km s$^{-1}$~}
\usepackage{comment}
\usepackage{orcidlink}

\received{}
\revised{}
\accepted{}
\submitjournal{ApJ}
\def \pl {\textit{Planck}~}

\shorttitle{Imprints of Stellar Feedback on Magnetic Fields in the Iris nebula NGC 7023}
\shortauthors{Sharma E. et al.}

\def\hii{H{\sc ii}}

\begin{document}

\title{Imprints of Stellar Feedback on Magnetic Fields in the Iris Nebula NGC 7023}

\email{sharma.ekta@bao.ac.cn, ektasharma.astro@gmail.com}

\author[0000-0002-4541-0607]{Ekta Sharma}
\affiliation{National Astronomical Observatories, Chinese Academy of Sciences, A20 Datun Road, Chaoyang District, Beijing 100012, People’s Republic of China}
\affiliation{Physical Research Laboratory, Navrangpura, Ahmedabad, Gujarat 380009, India}

\author[0000-0002-8557-3582]{Kate Pattle}
\affiliation{Department of Physics and Astronomy, University College London, Gower Street, London WC1E 6BT, United Kingdom}

\author[0000-0003-3010-7661]{Di Li}
\affiliation{New Cornerstone Science Laboratory, Department of Astronomy, Tsinghua University, Beijing 100084, China}
\affiliation{National Astronomical Observatories, Chinese Academy of Sciences, A20 Datun Road, Chaoyang District, Beijing 100012, People’s Republic of China}

\author[0000-0002-3179-6334]{Chang Won Lee}
\affiliation{Korea Astronomy \& Space Science Institute, 776 Daedeokdae-ro, Yuseong-gu, Daejeon, Republic of Korea}
\affiliation{University of Science and Technology, Korea (UST), 217 Gajeong-ro, Yuseong-gu, Daejeon 34113, Republic of Korea}

\author[0009-0007-0745-9147]{Maheswar Gopinathan}
\affiliation{Indian Institute of Astrophysics, Koramangala II Block, Bangalore 560 034, India}

\author[0000-0001-8516-2532]{Tao-Chung Ching}
\affiliation{National Radio Astronomy Observatory, 1003 Lopezville Road, Socorro, NM 87801, USA}

\author[0000-0001-8749-1436]{Mehrnoosh Tahani}
\affiliation{Kavli Institute for Particle Astrophysics and Cosmology (KIPAC), Stanford University, Stanford, CA, 51 United States}

\author[0000-0001-9333-5608]{Shinyoung Kim}
\affiliation{Korea Astronomy \& Space Science Institute, 776 Daedeokdae-ro, Yuseong-gu, Daejeon, Republic of Korea}



\begin{abstract}
We present 850 $\mu$m polarized continuum observations carried out with the POL-2 polarimeter mounted on the James Clerk Maxwell Telescope (JCMT) towards NGC 7023 located in the Cepheus Flare region. NGC 7023 is a reflection nebula powered by a Herbig Ae Be star HD 200775 and also identified as a \textit{hub} in the hub-filament cloud, LDN 1172/1174. We detect submillimetre emission well towards the northern (identified as C1) and the eastern region of the reflection nebula. We investigated the polarization structure and the magnetic field (B-field) morphology, which is found to be curved and follows the clump morphology. The comparison of the B-field morphology at the clump scales ($\sim$0.02 pc) with that of the envelope scale ($\sim$0.5 pc) suggests that the field lines are not preserved from envelope to clump scales, implying that an external factor may be responsible for disturbing the B-field structure. We estimated a magnetic field strength of 179$\pm$50 $\mu$G in the starless core, 121$\pm$34 $\mu$G in the protostellar core with a class I source and 150$\pm$42 $\mu$G in the protostellar core with a class II source using the \nthp (1$\textendash$0) line observed with 13.7 m single dish radio telescope at Taeduk Radio Astronomy Observatory (TRAO). The stability analysis using these B-field strengths gives magnetically sub-critical values, while the magnetic, gravitational, and outflow kinetic energies are roughly balanced. We also suggest that the reordering of the magnetic field lines could be due to the interaction with the already evolved high-velocity outflow gas around the central star, which hints at the presence of outflow feedback.
\end{abstract}



\section{Introduction} \label{sec:intro}
Magnetic fields (B-fields) play an important role in a broad range of astrophysical processes, ranging from regulating the star formation rate in the Milky Way \citep{Krumholz2019} to the evolution of galaxies \citep{Beck2015}. Given their omnipresence across all the scales of the interstellar medium (ISM), the magnetic fields affect different stages of star formation \citep{2004Ap&SS.292..225C,2023ASPC..534..193P}, i.e., the formation of filamentary structures \citep{2013ApJ...774..128S}, their fragmentation into cores \citep{Ching2022}, and in supporting the cores against the gravitational collapse \citep{2000ApJ...537L.135W,2022MNRAS.517.1138S}. The magnetic fields follow a parallel or perpendicular trend along or across the filamentary structures at different column densities \citep{2016A&A...586A.135P, Sharma2020}.

Massive stars ({\it M} $>$ 8 M$_{\odot}$) photoionize the surrounding gas by creating ionized atomic hydrogen regions (\hii) \citep{2007ARA&A..45..481Z}, which expand in the medium by creating shocks \citep{2018ARA&A..56...41M}. The ionized regions are responsible for the triggered star-formation through collect-and-collapse \citep{1977ApJ...214..725E} or radiative-driven implosion (RDI) process \citep{1989ApJ...346..735B}. 
Intermediate mass stars (2 $\leq$ {\it M}/M$_{\odot}$ $\leq$ 8) are lower mass counterparts of massive stars and due to less luminosities (5 $\leq$ {\it L$_{bol}$}/L$_{\odot}$ $\leq$ 10$^{4}$; \citealt{1993ApJ...418..414P}), they cannot create extended \hii~regions but drive feedback through the formation of photodissociated regions (PDRs). They are pre-main sequence stars, mainly Herbig Be or T-Tauri stars, and offer simpler physical environments than high-mass stars. The magnetic fields near \hii~regions have been characterized well and are found to exhibit diverse morphologies, e.g., complex in Monoceros R2 \citep{Hwang2022}, reordered in NGC 6334 \citep{Tahani2023} or unaltered by the stellar feedback in $\rho$ Oph A \citep{Kwon2018,2024A&A...690A.191L}. However, the magnetic fields around intermediate-mass stars are poorly constrained because only a handful of studies have covered the magnetic field strengths of the clouds in the vicinity of PDRs \citep{Pattle2018, Hwang2023}.

Three-dimensional radiation magnetohydrodynamic simulations for the expansion of the ionized regions in a uniform magnetized medium suggested that the B-field morphology in the surrounding natal cloud is compressed by the radiation from young stars \citep{2011MNRAS.414.1747A,2011MNRAS.412.2079M}. As a result, the field lines are aligned parallel to the neutral gas layers due to their interaction with the \hii~regions or PDRs. This finding agrees well with the several observational attempts at different extinction layers of the clouds, at smaller wavelengths \citep{2007ApJ...662.1014P,2017MNRAS.465..559S,2022RAA....22g5017C} and at longer wavelengths \citep{pattle2019a,2020ApJ...897...90E} with the magnetic fields being aligned with ionization fronts. 

Interstellar aspherical dust grains in the ISM are preferentially aligned with their major axes perpendicular to the local magnetic field orientation projected onto the plane-of-sky \citep{1953ApJ...118..113C,1951ApJ...114..206D,2007MNRAS.378..910L}. Hence, the thermal emission from dust grains is linearly polarized and can be used as an efficient tool to probe the magnetic fields of molecular clouds at submillimeter/millimetre wavelengths \citep{2012ARA&A..50...29C}. We have employed submillimeter polarization in this work to investigate the role of magnetic fields at clump scales in the vicinity of stars.

Nearby reflection nebulae are the best indicators of recent or ongoing star-formation activity as the radiation field around the massive star heats the surrounding gas and modifies the chemistry of structures, making them illuminating \citep{2003A&A...399..141M}. We chose NGC 7023, a reflection nebula illuminated by an intermediate-mass star. This nebula offers a simpler environment in order to investigate the dynamical importance of the magnetic fields toward dense cloud structures under stellar feedback.

\section{Target Selection: NGC 7023}
\begin{figure*}[ht!]
\includegraphics[height=7cm, width=12cm,trim={0 0 0 0}]{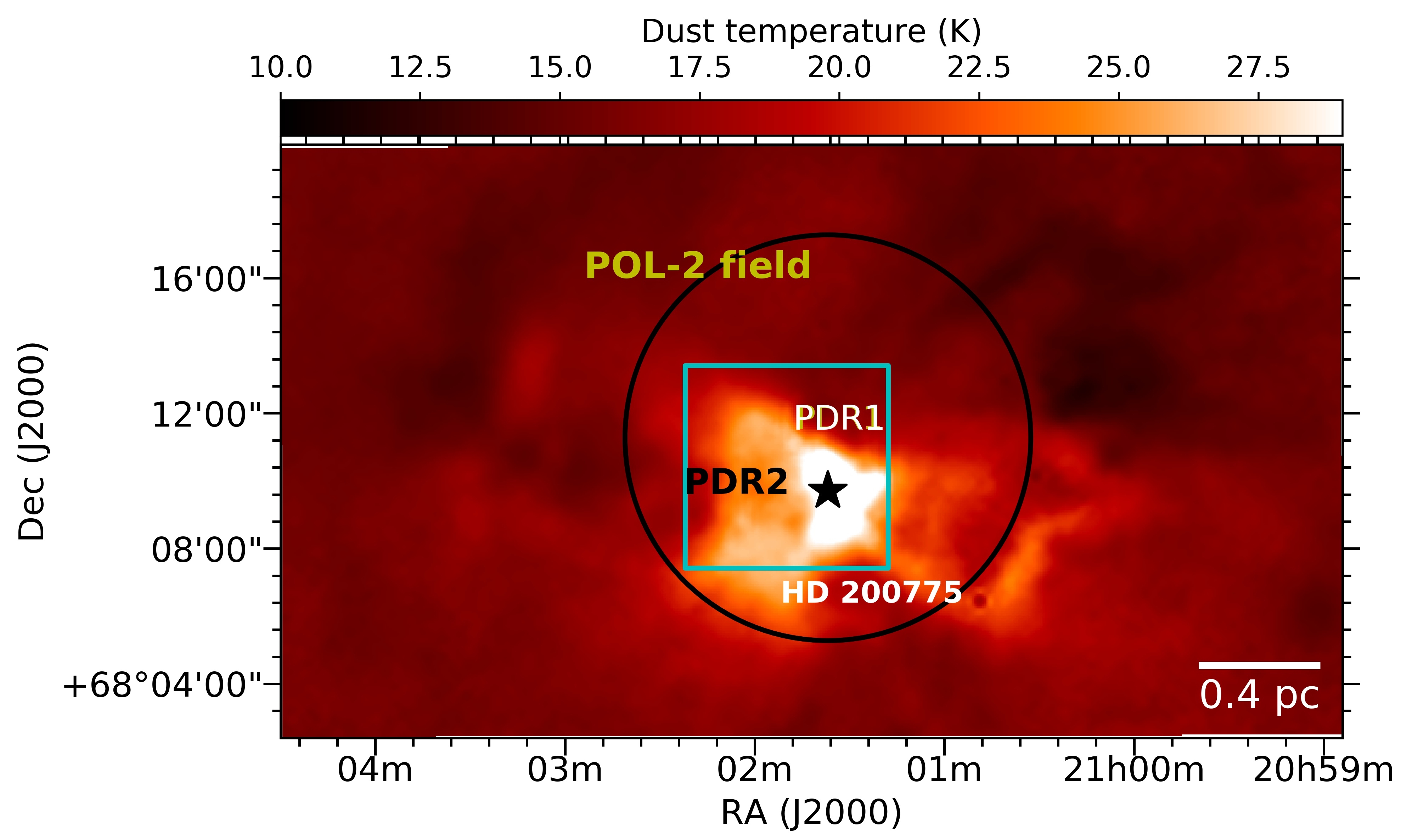}
\includegraphics[height=6cm, width=7cm,trim={1cm 0 0 2cm}]{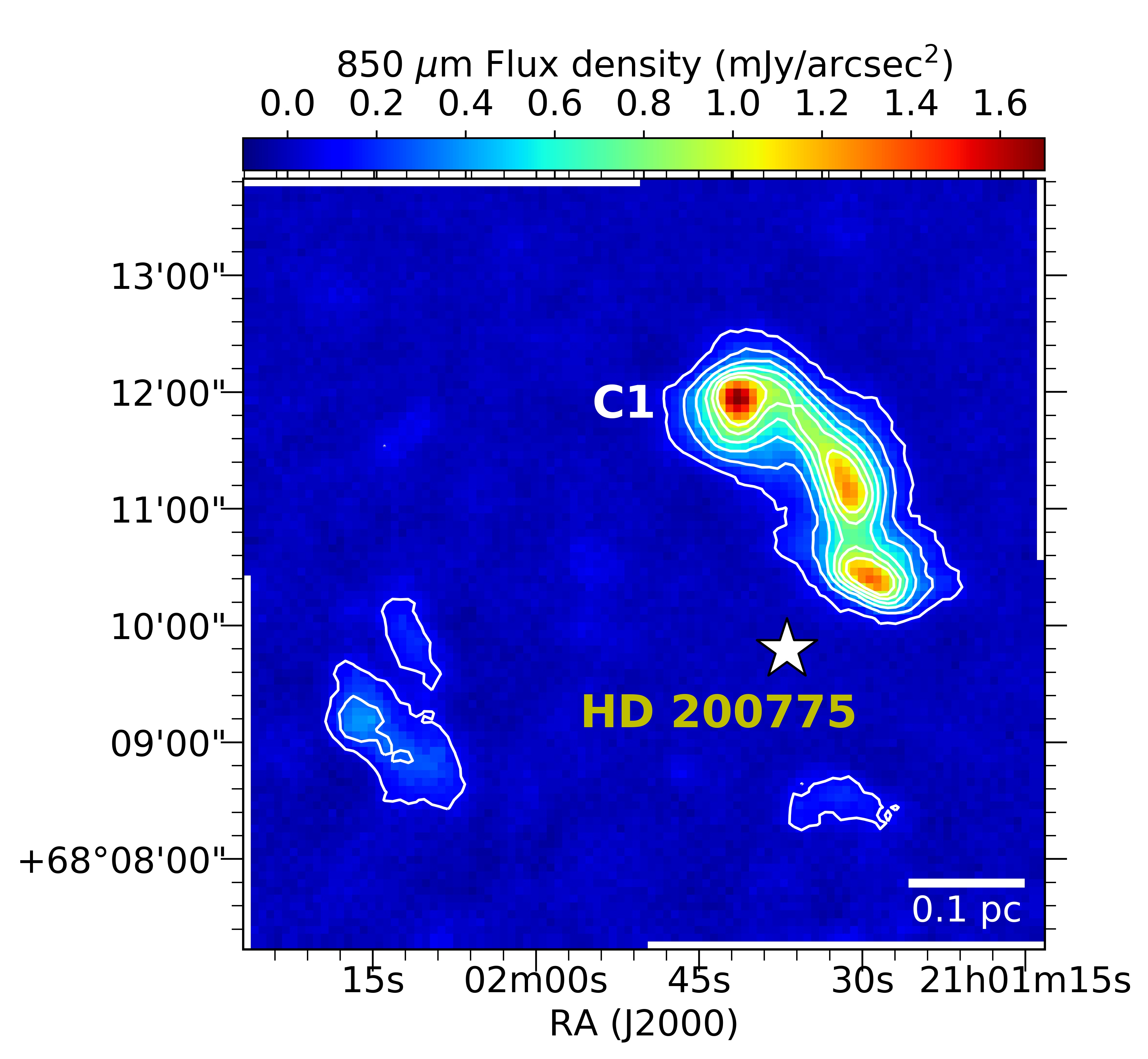} 
\caption{{\bf Left:} Dust temperature map of NGC 7023 derived from \textit{Herschel} maps of L1172/1174 cloud \citep{2020ApJ...904..172D}. The central star is marked in black with PDRs labelled. The solid black circle marks the POL-2 field of 6$^{'}$ radius. The cyan box shows the area of interest considered in later figures. {\bf Right:} SCUBA-2 850 $\mu$m Stokes I map obtained with our polarization observations. The clump C1 is labelled.}\label{fig:stokes_I}
\end{figure*}

The NGC 7023 (the \textit{Iris} nebula) [RA: 21h 01m 37s, Dec: +68$^{\circ}$ 09$^{'}$ 48$^{''}$ (J2000)] is a reflection nebula located in the Cepheus Flare constellation at the distance of 340 pc \citep{2020MNRAS.494.5851S}. The spatial extent of the nebula in the plane-of-sky is 1.5 pc (angular extent of $\sim$15$^{'}$ in the east-west direction). The nebula hosts an intermediate-mass star HD 200775, which is a spectroscopic binary system of spectral type B3Ve-B5 \citep{2008MNRAS.385..391A}. The region is termed a \lq\lq{hub-filament}\rq\rq complex by \cite{2009ApJ...700.1609M} with NGC 7023 sitting at the location of the hub. The region holds the highest concentration of young stellar objects (YSOs) in all Cepheus Flare clouds \citep{2009ApJS..185..198K}. This region contains PDRs, high-density regions, and diffuse gas streamers seen in gas and dust emission maps. 
For the present study, we chose this particular region of the nebula because it is in the closest proximity to the central star and is expected to show possible signatures of stellar feedback. The magnetic field structure of NGC 7023 has been studied using starlight polarization at $\sim$ 0.6 $\mu$m using the Aries IMaging POLarimeter (AIMPOL) by \cite{Saha2021}. The global magnetic fields are quite regular. With the high sensitivity of POL-2, we mapped the magnetic fields in this region, which has present feedback from an intermediate-mass star (see the solid black circle marked in Fig. \ref{fig:stokes_I} (left panel).

The paper is organized as follows. In Section 3, we describe our observations and the data reduction of the JCMT POL-2 and the molecular line observations using the Taeduk Radio Astronomy Observatory
(TRAO). In Section 4, we present the results of our 850 $\mu$m continuum polarization observations and the resulting magnetic field morphology towards the reflection nebula. We study the dust grain properties using the Ricean model on the obtained polarization fractions and also derive the magnetic field strength using the Davis-Chandrasekhar-Fermi method. In Section 5, we finally discuss our results and compare the magnetic field energy with the turbulent kinetic, gravitational, and outflow energies. Based on the results, we suggest an evolution scenario of the clump around PDR. We finally summarize our results in Section 6.

\section{Observations and data reduction}
\subsection{JCMT POL-2 data}
We observed the NGC 7023 region 34 times between 2021 August 07 and 2022 May 24 using the POL-2 polarimeter \citep{friberg2016} mounted on the Submillimetre Common-User Bolometer Array 2 (SCUBA-2; \citealt{holland2013}) on the James Clerk Maxwell Telescope (JCMT). The data were taken in a mixture of Band 1 ($\tau_{225\,{\rm GHz}}<0.05$) and Band 2 ($0.05<\tau_{225\,{\rm GHz}}<0.08$) weather under project codes M21BP004 and M22AP012. Each observation consisted of a 31-minute POL-2-DAISY scan pattern.

The data were reduced using the $pol2map$\footnote{\url{http://starlink.eao.hawaii.edu/docs/sun258.htx/sun258ss73.html}} script recently added to the \textsc{Smurf} package in the $Starlink$ software suite \citep{chapin2013}. See \citet{pattle2021a} for a detailed description of the current POL-2 data reduction process. Instrumental polarization (IP) was corrected for using the `August 2019' IP model\footnote{\url{https://www.eaobservatory.org/jcmt/2019/08/new-ip-models-for-pol2-data/}}. The 850 $\mu$m data were calibrated with a flux conversion factor (FCF) of 2795 mJy\,arcsec$^{-2}$\,pW$^{-1}$ using the post-2018 June 30 SCUBA-2 FCF of 2070 mJy\,arcsec$^{-2}$\,pW$^{-1}$ \citep{mairs2021} multiplied by a factor of 1.35 to account for additional losses in POL-2 \citep{friberg2016}. POL-2 observes simultaneously at 850 $\mu$m and 450 $\mu$m, but we consider only the 850 $\mu$m observations in this work as the 450 $\mu$m data reduction process remains under development.

We binned our output vector catalogue to 8-arcsec (approximately Nyquist-sampled) pixels.
The average RMS noise in Stokes $Q$, $U$, and $I$ in the central 3 arcmin of the map on 8-arcsec pixels is 0.8 mJy\,arcsec$^{-2}$.

The observed polarised intensity is given by
\begin{equation}
    PI^{\prime} = \sqrt{Q^{2} + U^{2}}.
\end{equation}
We debiased this quantity using the modified asymptotic estimator \citep{plaszczynski2014,montier2015}:
\begin{equation}
    PI = PI^{\prime} - \frac{1}{2}\frac{\sigma^{2}}{PI^{\prime}}\left(1-e^{-\left(\frac{PI^{\prime}}{\sigma}\right)^{2}}\right),
\end{equation}
where $\sigma^{2}$ is the weighted mean of the variances {$\sigma_{Q}^{2}$ and $\sigma_{U}^{2}$},
\begin{equation}
    \sigma^{2} = \frac{Q^{2}\sigma_{Q}^{2} + U^{2}\sigma_{U}^{2}}{Q^{2} + U^{2}},
\end{equation}
calculated on a pixel-by-pixel basis. The debiased polarization fraction is given by $p = PI/I$.

The polarization angle is given by
\begin{equation}
    \theta_{p} = 0.5\arctan(U,Q),
\end{equation}
We note that the polarization angles we detect are not true vectors, as they occupy a range in angle $0-180^{\circ}$. We nonethelesss refer to our polarization angle measurements as vectors for convenience to keep consistency with the general convention used in this field.
Throughout this work, we assume that dust grains are aligned with their major axis perpendicular to the magnetic field direction \citep[e.g.][]{andersson2015}, and so that the plane-of-sky magnetic field direction can be inferred by rotating $\theta_{p}$ by 90$^{\circ}$.

\subsection{Molecular line data}
We used molecular line data of C$^{18}$O (1$\textendash$0), N$_{2}$H$^{+}$ (1$\textendash$0), and CS (2$\textendash$1) observed using 14 m single dish telescope at Taeduk Radio Astronomy Observatory (TRAO), South Korea from November-December 2018. The observations were carried out as a part of mapping the large cloud L1172/1174 using four tracers. We used the new receiver system Second QUabbin Observatory Imaging Array-TRAO (SEQUOIA-TRAO) in the 85-115.6 GHz frequency range. The pointing accuracy was achieved to be $\leq$ $5^{''}$ using a standard X Cygnus source in the SiO line. The back-end system with a Fast Fourier transform spectrometer has 4096 $\times$ 2 channels at 15 kHz resolution ($\sim$0.05 \kms at 110 GHz). The spatial resolution was 47$^{''}$ at 109.7821734 GHz, and the final velocity resolution for the data was adjusted to be 0.1 \kms. The system temperature ({\it T$_{sys}$}) was 500-700 K during the observations. The typical rms noise ({\it T$_{A}^{\ast}$}) in a channel of 0.1 \kms is $\sim$0.09 K for \coo and $\sim$0.06 K for \nthp (1$\textendash$0) and 0.1 K for CS (2$\textendash$1) line in a channel of 0.06 \kms. The rest frequencies used in our observations for CS (2$\textendash$1) and \nthp (JF1F = 101 - 012) are 96.412953 GHz and 93.176258 GHz, respectively \citep{2001ApJS..136..703L}. We used the astronomical software package Gildas/CLASS\footnote{\url{http://www.iram.fr/IRAMFR/GILDAS/}} for data reduction. 

The \coo line (critical density, n$_{crit}\sim$10$^{3}$ - 10$^{4}$ cm$^{-3}$) traces dense gas in the filaments while the \nthp (1$\textendash$0) line (n$_{crit}$ $\sim$ 10$^{5}$cm$^{-3}$; \citealt{2015PASP..127..299S}) traces relatively denser gas, particularly in the prestellar phase \citep{2002ApJ...572..238C,2016A&A...587A.118P}. The CS (2$\textendash$1) is also a dense gas tracer (n$_{crit}\sim$ 1.3$\times$10$^{5}$ cm$^{-3}$; \citealt{2015PASP..127..299S}), particularly chosen to study infall motions in prestellar cores \citep{2001ApJS..136..703L}. These three molecular lines are typical gas tracers used to probe the kinematics and physical conditions of the dense gas in star-forming clumps. In order to study the variations of the magnetic field strength around cloud and dense core regions, we have used different gas tracers in further analysis.

\begin{figure*}[ht!]
\centering
\includegraphics[height=7cm, width=9cm]{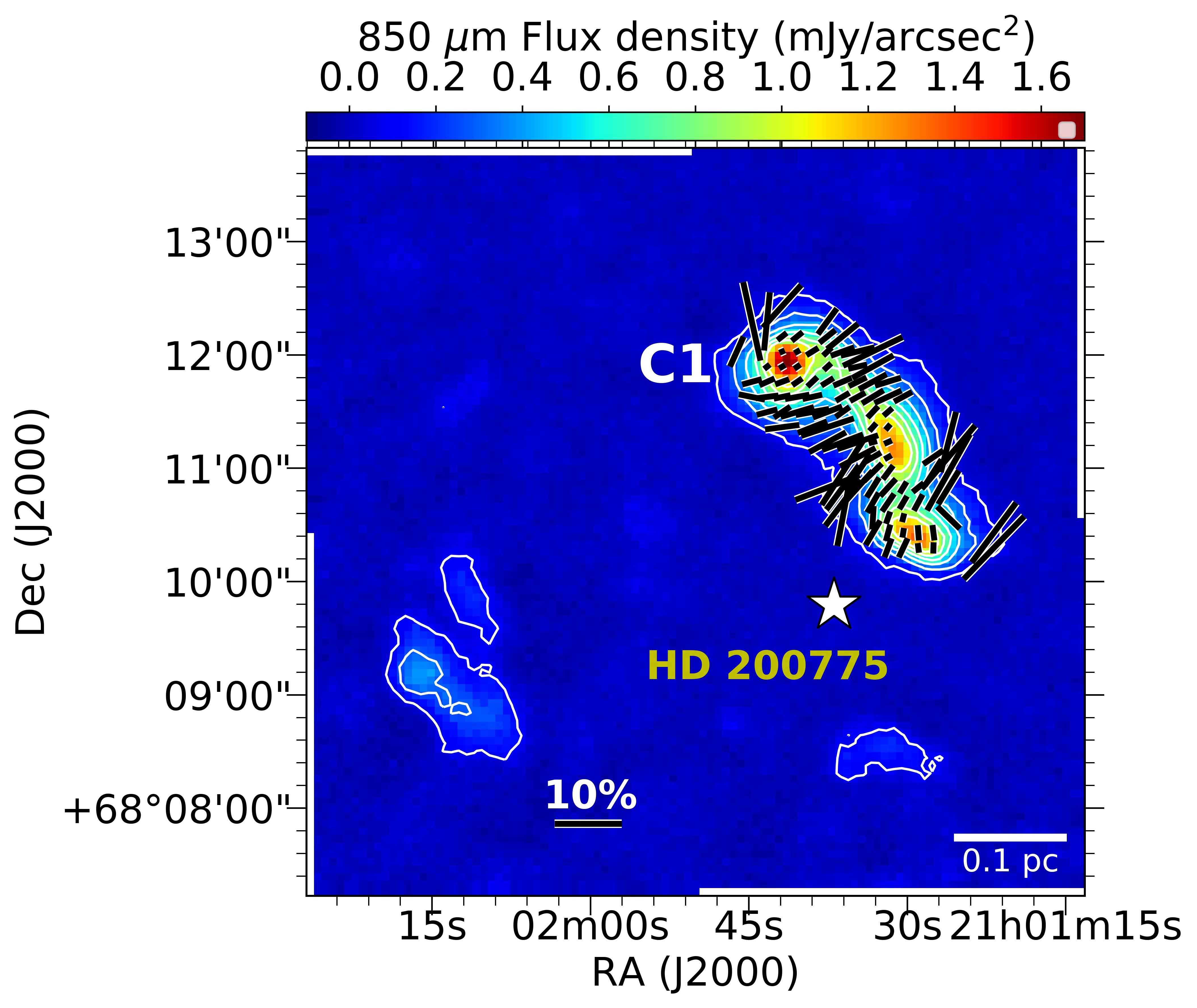} 
\includegraphics[height=7cm, width=8.5cm]{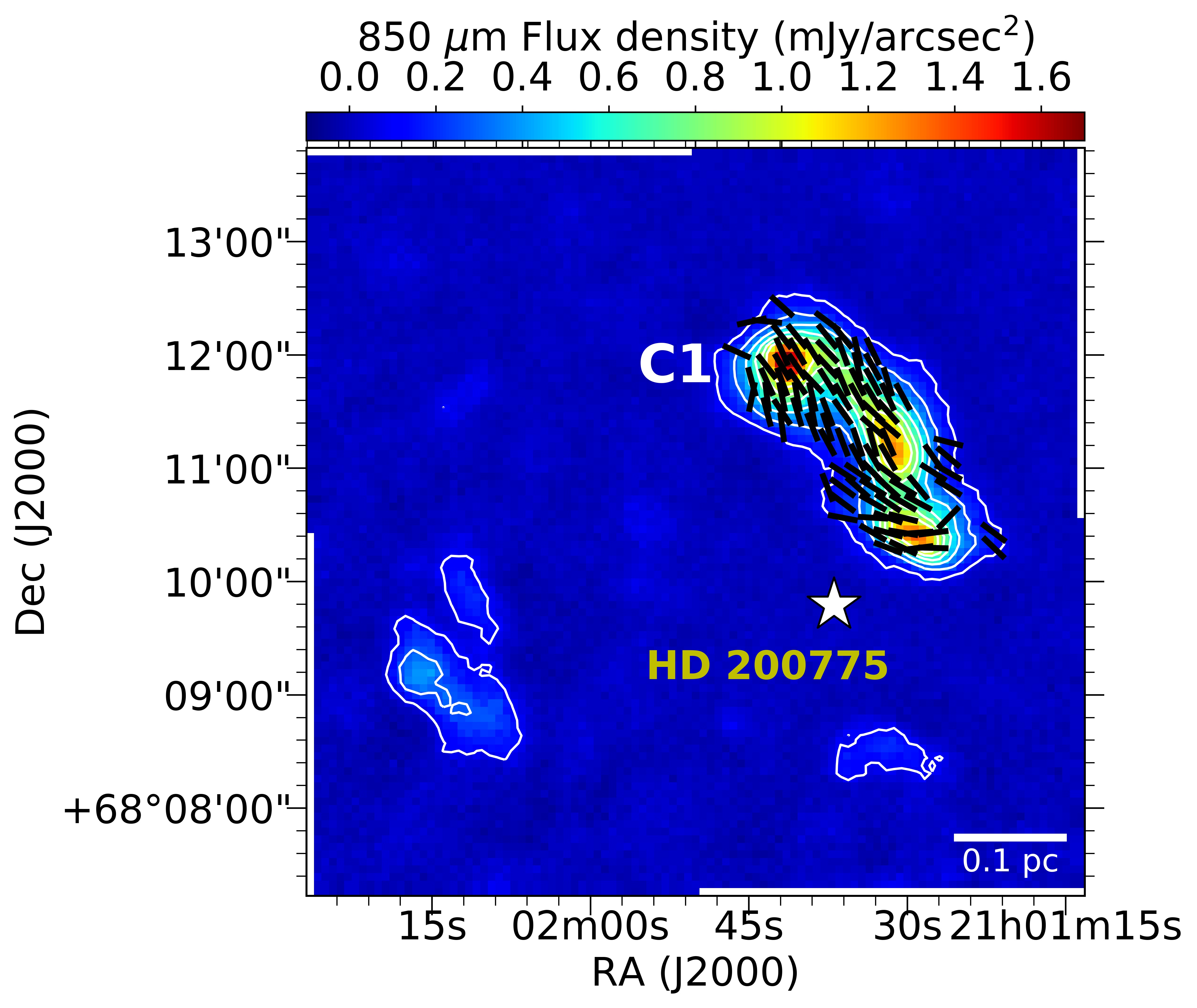}
\caption{{\bf Left:} The POL-2 dust polarization map at 850 $\mu$m towards NGC 7023 reflection nebula near the Herbig Ae Be star, HD 200775. The background image shows the Stokes I intensity. The black lines show the polarization angles and the degree of polarization in their length is proportional to polarization percentages. The reference vector with a polarization percentage of 10$\%$ is shown. {\bf Right:} Magnetic field map towards NGC 7023 with equal length vectors which have been rotated by 90$^{\circ}$ to trace the plane-of-sky magnetic field direction.}\label{fig:1_pol}
\end{figure*}

\section{Results} \label{sec:highlight}
\subsection{Magnetic field morphology}
Fig. \ref{fig:stokes_I} shows the dust temperature map of NGC 7023 on the left and the detected 850 $\mu$m Stokes I map obtained from SCUBA-2 observations of the observed region towards NGC 7023 on the right. The location of PDRs is labelled. Investigating the structure and physical conditions of the PDRs is important for understanding the effect of stellar feedback. The outflow has carved out a biconical cavity and created the PDRs, marked as PDR 1 and PDR 2 \citep{2014A&A...569A.109K} triggered as a result of HD 200775 (B3Ve-B5 spectral type). The solid black circle marks the region observed in submillimetre polarization using POL-2 mounted on JCMT, and the cyan box represents the region of interest covered in further analysis. The right panel shows the obtained SCUBA-2 dust emission map, and the white contours are drawn at the emission level higher than 3$\sigma$, which is mainly present towards three regions, one towards the north (identified as clump C1) and a small extent towards the east and the south. The detected emission correlates with the position of marked PDRs, which are present towards dense gas structures' rims. 

Fig. \ref{fig:1_pol} (left) shows the distribution of polarization angles towards the clump C1. The vectors were selected on the basis of p$_{db}/\delta$p $>$ 3, I/$\delta$I $>$ 5, and $\delta\theta<$ 10$^{\circ}$, where the last two conditions are degenerate considering the Serkowski's approximation, $\delta\theta$ $\approx$ 28.65 $\times$ $\delta$p/p \citep{1962AdA&A...1..289S}. The vectors in the regions to the east and south of the star HD 200775 do not satisfy the selection criteria.
The polarization fraction shows a decrease towards the high-intensity regions. This depolarization may result from a combination of field tangling and the loss of grain alignment at elevated optical extinctions \citep{andersson2015}. At lower visual extinctions (A$_{V}$), dust grains tend to align their minor axis along the direction of the local magnetic field \citep{1951ApJ...114..206D,whittet2008,2014A&A...569L...1A}. However, as the optical depth increases, dust grains tend to lose their alignment due to the absence of a non-isotropic radiation field, which is crucial in the spinning of asymmetrical dust grains in accordance with the Radiative Alignment Torque (RAT) theory of grain alignment \citep{2007MNRAS.378..910L}. We quantify the efficiency of grain alignment in more detail in section 4.4. In the right panel of Fig. \ref{fig:1_pol}, we show the distributions of magnetic field orientation (black lines), which are obtained by rotating the polarization angles by 90$^{\circ}$. The B-fields exhibit a preferred orientation of 35$^{\circ}$ E of N and align with the clump morphology. In the next section, we further quantify the correlation of clump morphology with the B-field orientation.

\subsection{Comparison of filament and magnetic field orientation}
\begin{figure}[ht!]
    \centering
\includegraphics[width=9cm, height=7.5cm]{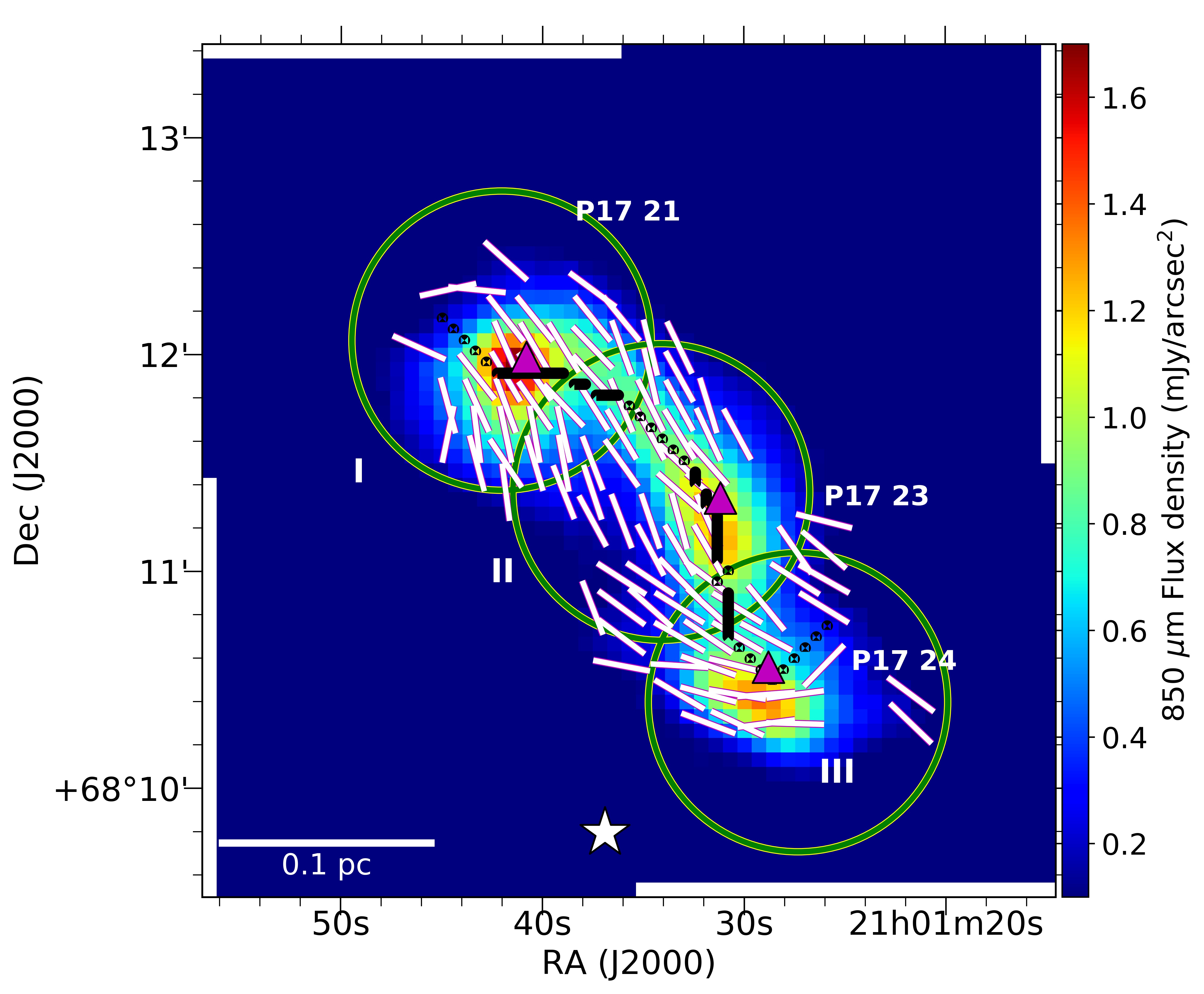}
    \caption{Extracted filament in black dotted line overplotted on the SCUBA-2 image with the magnetic field lines shown in white. The magenta triangles show three cores labelled as P17 21, P17 23 and P17 24. The names were adopted from \cite{pattle2017}. The circles show the regions I, II, and III, chosen to calculate the average magnetic field around the three cores.}
    \label{fig:fil_ngc}
\end{figure}
As mentioned in the previous section, the magnetic field orientation inferred from POL-2 data is curved and runs along the clump morphology. This clump shows a high aspect ratio and hosts three cores, which are protostellar and starless \citep{pattle2017}. Therefore, we considered this a filament and extracted the filament orientation using \textit{Filfinder} algorithm \citep{2015MNRAS.452.3435K}. We used a \textit{global threshold} of 0.6 mJy/arcsec$^{2}$ and the \textit{adaptive threshold} as 8 pixels on the SCUBA-2 map. We used \textit{size threshold} of 200 square pixels to extract filaments with lengths down to 0.3 pc. Fig. \ref{fig:fil_ngc} shows the high-density filament in black derived using \textit{Filfinder} algorithm. Three high-density cores are marked towards the clump C1 in 850 $\mu$m emission (magenta triangles). The dust emission properties of these cores have been studied by \cite{pattle2017}, and hereafter, we adopt their names P17 21 for starless, P17 23 for protostellar with class I source and P17 24 for protostellar with class II source. Since the region shows curved morphology, we examined the polarization angles of vectors that are nearest neighbours to the filament. Further, we divided the filament into three regions - I, II and III, to compare the orientation of the filament segment in each region and the average B-field orientation. The regions are shown by solid green circles in Fig. \ref{fig:fil_ngc}. The average magnetic field within the I region, $\langle${$\theta_{B}^{I}$}$\rangle$ = 26.4$\pm$26.0$^{\circ}$, and the orientation of the filament segment, $\theta_{f}^{I}$=24$^{\circ}$ E of N. The average magnetic field in region II, $\langle${$\theta_{B}^{II}$}$\rangle$ = 35$^{\circ}\pm$14$^{\circ}$, and the $\theta_{f}^{I}$ = 45$^{\circ}$, E of N. In region III, the $\langle${$\theta_{B}^{III}$}$\rangle$ = 39$^{\circ}\pm53^{\circ}$, and the $\theta_{f}^{III}$ is 63$^{\circ}$, E of N around P17 24 core. The orientation of the B-field and the filament matches well within the I and II regions. However, the dispersion of B-field orientations is higher in Region III as the filament segment exhibits more curvature. 

\begin{figure*}[hbt!]
\centering
\includegraphics[width=10cm, height=7cm]{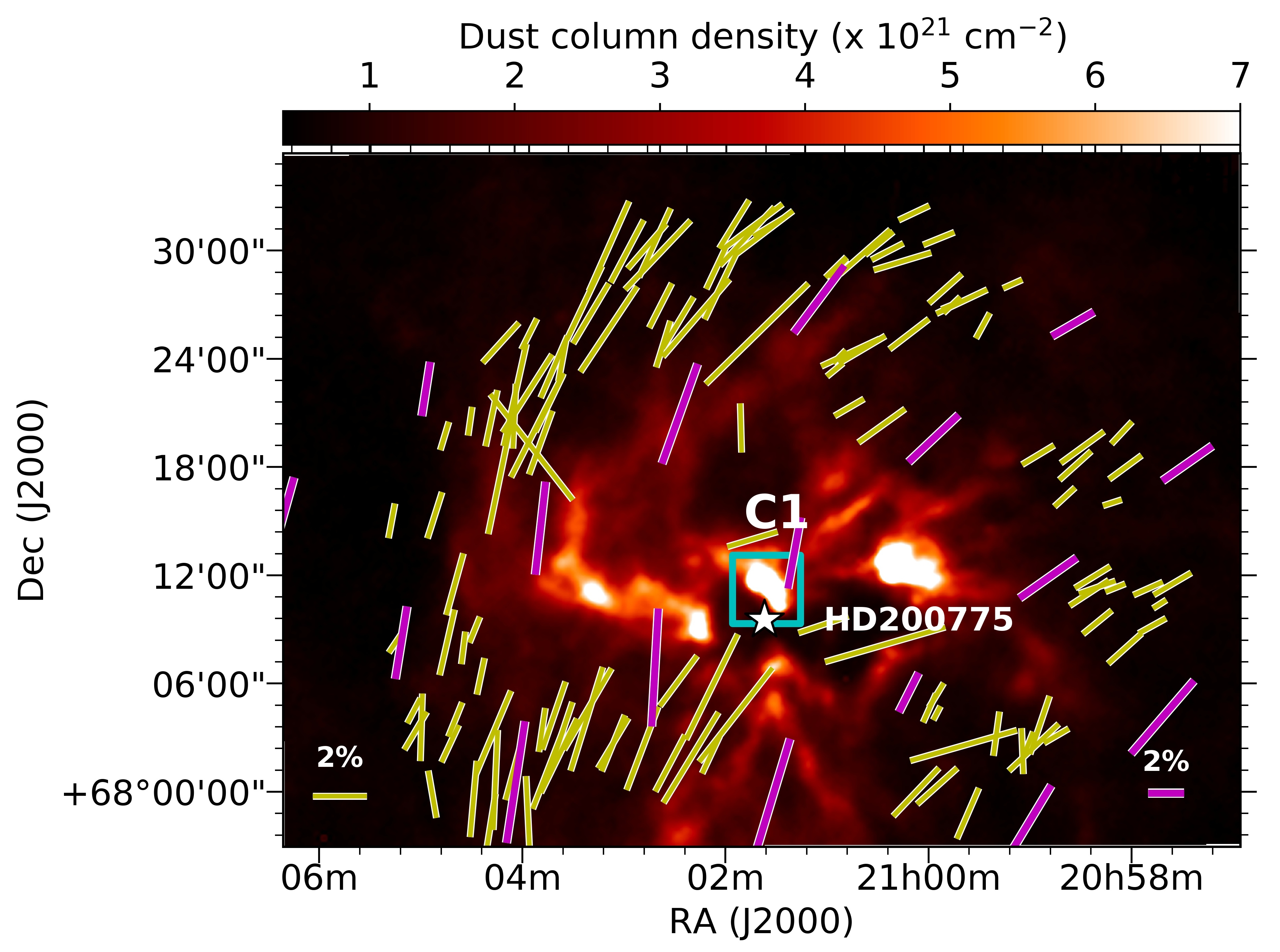}
\includegraphics[width=6cm, height=6cm, trim={0 0.9 0 0.3}]{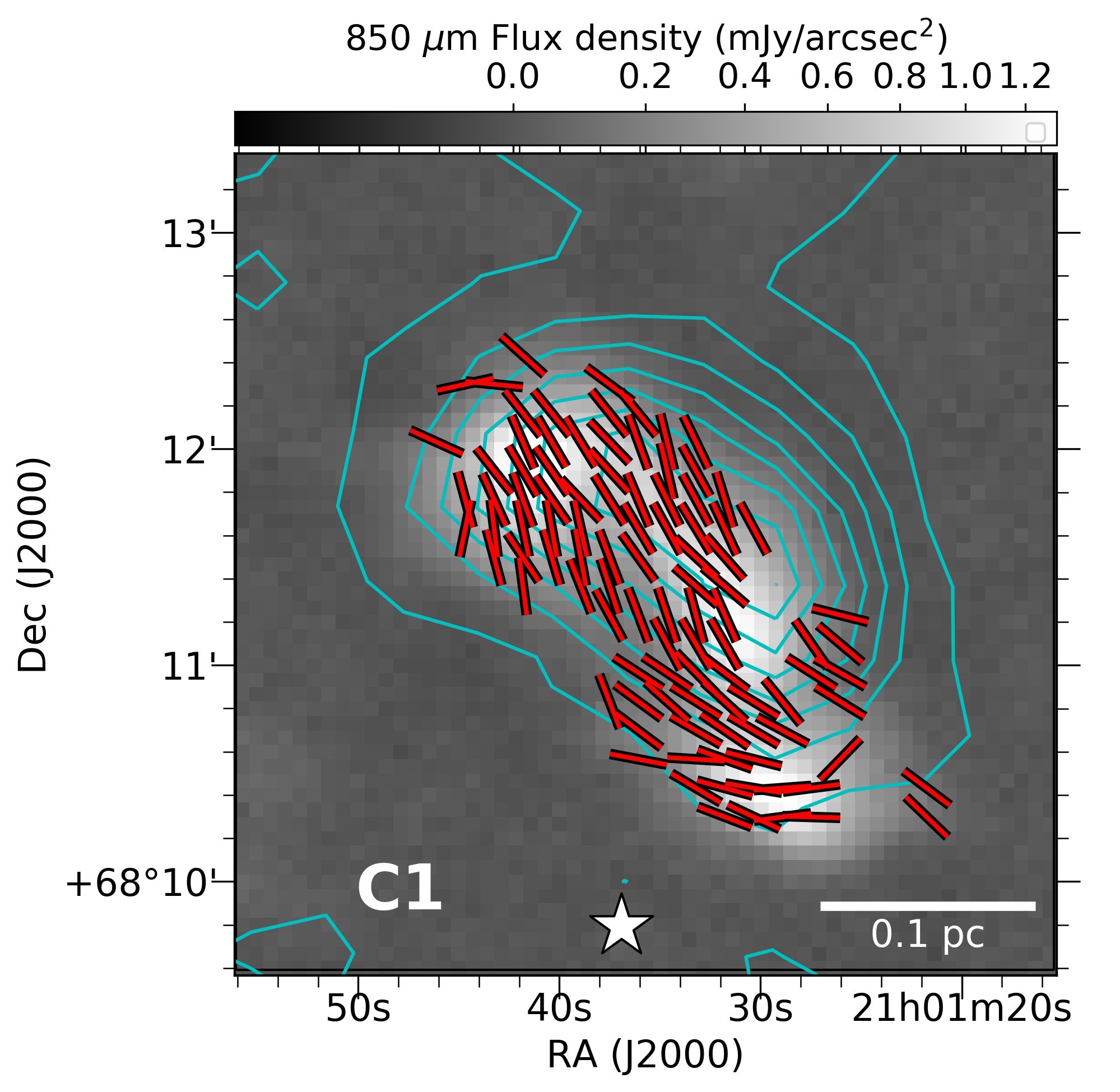}
\caption{{\bf Left:} \textit{Herschel} dust column density map of NGC 7023. The white star marks the position of the intermediate-mass star HD 200775. The magenta lines show the magnetic fields derived from \textit{Planck} polarization data, and the yellow lines show the magnetic field lines measured using optical polarization vectors \citep{Saha2021}. {\bf Right:} SCUBA-2 850 $\mu$m Stokes I map towards C1. The zoomed-in region is marked with a cyan box in the main image. The integrated intensity contours of N$_{2}$H$^{+}$ (1$\textendash$0) in cyan are in the range from 0.47-1.4 K kms$^{-1}$ (step size $\sim$ 0.1 K kms$^{-1}$). Red vectors show our POL-2 observations towards C1 with $p/dp > 3$.}\label{fig:pl_op} 
\end{figure*}

\subsection{Comparison with large-scale magnetic field structure from \textit{Planck} and starlight polarization}
\begin{figure}[hbt!]
\centering
\includegraphics[height=8cm, width=8.5cm]{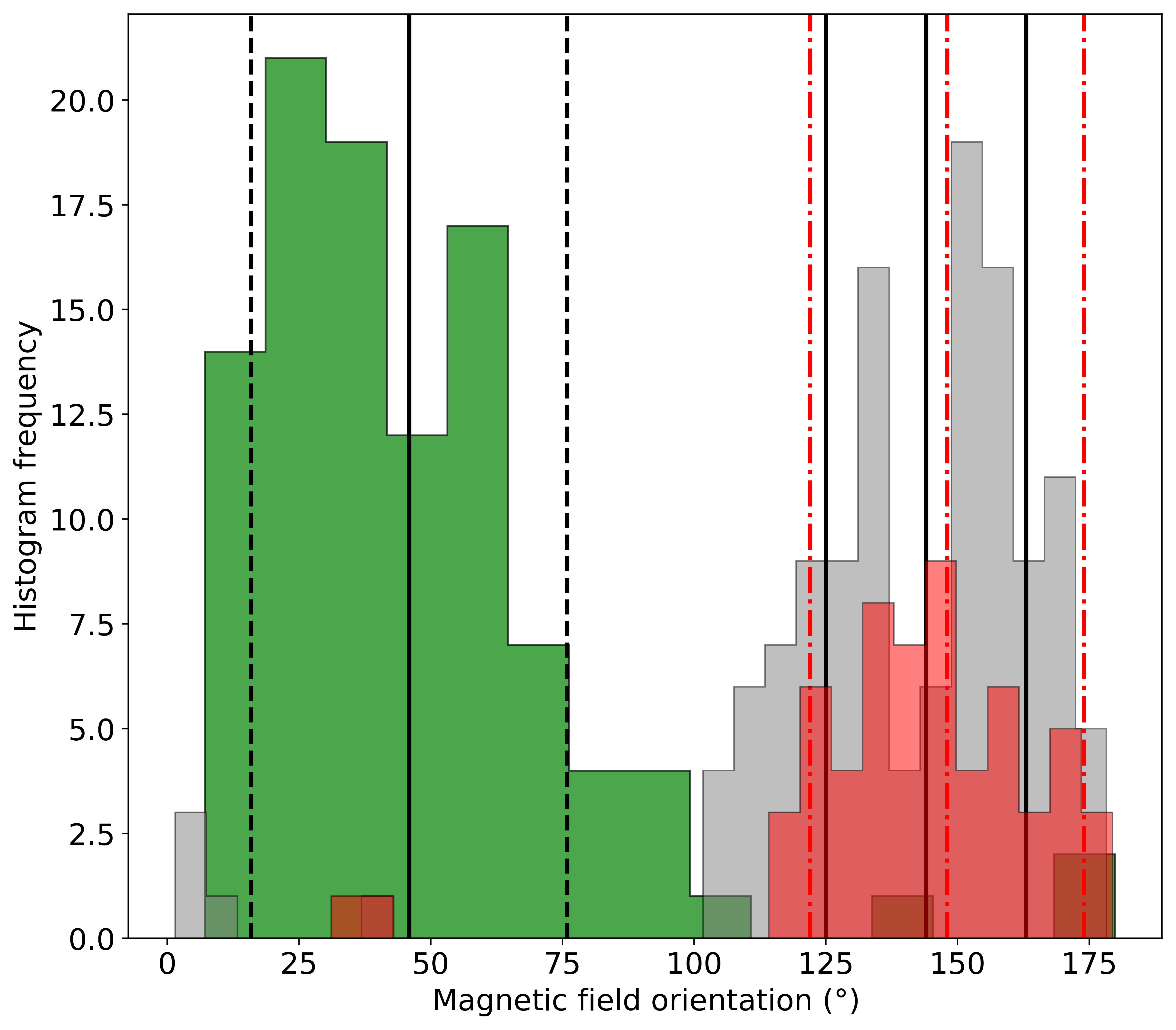} 
\caption{The distribution of polarization angles rotated by 90$^{\circ}$ to infer the magnetic field orientations obtained using JCMT POL-2 is shown in green. The B-field orientations from optical R-band \cite{Saha2021} and from \pl polarization data (rotated by 90$^{\circ}$) are plotted in grey and red, respectively. The black and red vertical lines around the R-band ($\sim$ 0.6 $\mu$m) and \pl distribution are the mean magnetic field direction with their range of dispersion.}\label{fig:hist}
\end{figure}

The \textit{Planck} polarization data at 353 GHz (850 $\mu$m) are used to study the large-scale magnetic fields around NGC 7023 at a resolution of 5$^{'}$ (0.5 pc) of the \pl beam. \cite{Saha2021} have studied the large-scale magnetic field using high-resolution starlight polarization measurements at optical wavelengths as well as the \textit{Planck} data. Fig. \ref{fig:pl_op} shows the magnetic field vectors from optical data (in yellow) and \textit{Planck} (in magenta) overplotted over the \textit{Herschel} column density image. In the zoomed version of the small inset marked with a cyan box on the left image, we show the POL-2 vectors (red lines) overplotted on the SCUBA-2 850 $\mu$m emission map. The optical polarization and the \pl polarization vectors agree well with each other, whereas the structure of the magnetic fields is quite different at clump scales. The diffuse regions on the western and eastern sides of HD 200775 show a fairly regular magnetic field with a northwest-southeast orientation following the diffuse structures. The same magnetic fields at larger scales are broadly perpendicular to the major axis of the high-density structure on the eastern side and parallel to that on the western side. This is similar to what we expect from the bimodal distribution of magnetic fields \citep{2013ApJ...774..128S,2016A&A...586A.138P}. In contrast, the POL-2 vectors appear reordered at smaller scales and follow the curvature of the clump. 
  
Fig. \ref{fig:hist} shows the distribution of polarization angles obtained with POL-2 in green colour. The mean polarization angle, $\langle{\theta_{POL2}}\rangle$= 46$\pm30^{\circ}$ E of N. The distribution of \textit{Planck} and optical polarization angles are shown in grey and red histograms, respectively. The mean values from both the data sets are $\langle{\theta_{Planck}}\rangle$= 148$^{\circ}\pm26^{\circ}$ and $\langle\theta_{optical}\rangle$ = 144$^{\circ}\pm$19$^{\circ}$ E of N. The magnetic field structure inferred from both \textit{Planck} and optical polarization follows a similar distribution. The dispersion of POL-2 data is broader, and notably, the morphology of the magnetic field traced by POL-2 is completely different from the large-scale field orientation, which is well aligned with the diffuse low-column density structures. This difference in orientation between large and small scales suggests that the cloud-scale B-field is not preserved in dense gas.

\subsection{Grain alignment properties}
Polarized dust emission observations typically show a power-law dependence, $p\propto I^{-\alpha}$, where $0\leq \alpha \leq 1$ \citep{whittet2008,jones2015}.  A steeper index (higher $\alpha$) indicates either poorer grain alignment with respect to the magnetic field or more variation of the magnetic field direction along the line of sight (LOS): $\alpha = 0$ indicates that grains are consistently aligned throughout the LOS, while $\alpha = 1$ implies complete randomization of either grain alignment or magnetic field direction along the LOS \citep{pattle2019a}.

We measured $\alpha$ using the method described by \citet{pattle2019a}, in which we assume that the underlying relationship between $p$ and $I$ can be parametrized as
\begin{equation}
    p = p_{\sigma_{QU}}\left(\frac{I}{\sigma_{QU}}\right)^{-\alpha},
    \label{eq:polfrac}
\end{equation}
where $p_{\sigma_{QU}}$ is the polarization fraction at the RMS noise level of the data $\sigma_{QU}$, and $\alpha$ is a power-law index in the range $0 \leq \alpha \leq 1$. We fitted the relationship between $I$ and observed non-debiased polarization fraction $p^{\prime}$ with the mean of the Ricean distribution of observed values of $p$ which would arise from equation~\ref{eq:polfrac} in the presence of Gaussian RMS noise $\sigma_{QU}$ in Stokes $Q$ and $U$:
\begin{equation}
    p^{\prime}(I) = \sqrt{\frac{\pi}{2}}\left(\frac{I}{\sigma_{QU}}\right)^{-1}\mathcal{L}_{\frac{1}{2}}\left(-\frac{p_{\sigma_{QU}}^{2}}{2}\left(\frac{I}{\sigma_{QU}}\right)^{2(1-\alpha)}\right).
    \label{eq:rmfit}
\end{equation}
where $\mathcal{L}_{\frac{1}{2}}$ is a Laguerre polynomial of order $\frac{1}{2}$. See \citet{pattle2019a} for a derivation of this result. We restricted our data set to the central 3-arcminute diameter region over which exposure time, and so RMS noise, is approximately constant \citep{friberg2016}.

\begin{figure}
    \centering
    \includegraphics[width=0.47\textwidth]{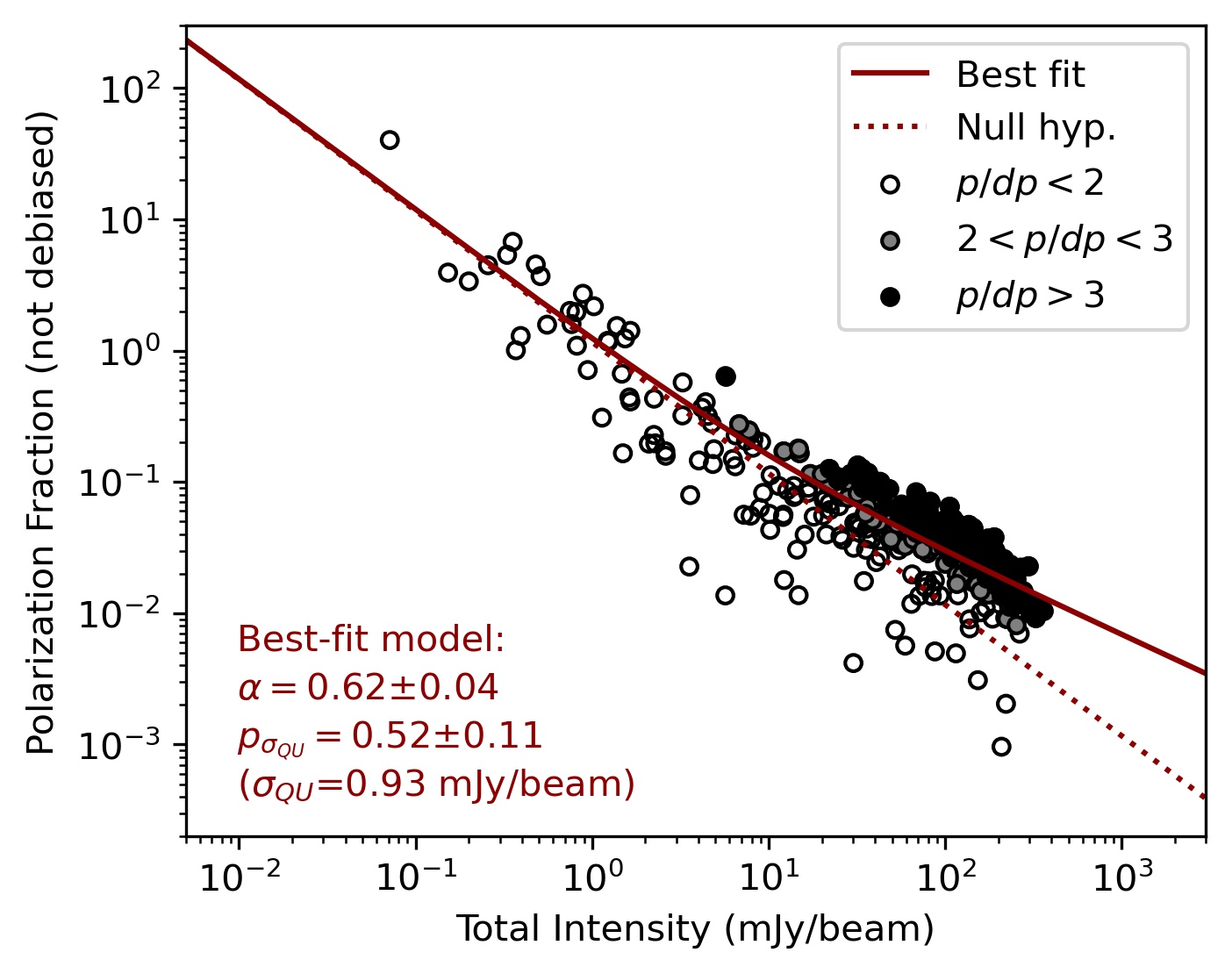}
    \caption{Non-debiased polarization fraction as a function of Stokes $I$ intensity, fitted with a single-power-law distribution and a Ricean noise model, as described in the text. All data points in the central 3-arcmin-diameter region of the image are shown and fitted; those at $p/dp > 2$ are shaded in grey; those at $p/dp > 3$ are shaded in black. The best-fit model, with a power-law index $\alpha=0.62\pm0.04$, is shown as a solid maroon line. The behaviour in the absence of true polarised signal, $p^{\prime} = \sqrt{\pi/2}(I/\sigma_{QU})^{-1}$, is shown as a dashed maroon line.}
    \label{fig:p_vs_I}
\end{figure}

The relationship between $p^{\prime}$ and $I$ in NGC 7023 is shown in Figure~\ref{fig:p_vs_I}.  By fitting Equation~\ref{eq:rmfit} to the data, we measure a best-fit index of $\alpha = 0.62\pm 0.04$. This suggests that in our observations of NGC 7023, grains remain fairly well-aligned even at the highest column densities.
A power law index of 0.5 implies a linear decrease in the polarization fraction, suggesting that the dust grain alignment decreases linearly with the depth into the cloud. This trend of decreasing polarization as a function of increasing Intensity has been seen in other molecular clouds also \citep{2018ApJ...861...65S, Kwon2018,pattle2019a,2022ApJ...926..163K}. This could result from the decreasing radiation fields and increasing gas density or larger grain sizes in dense regions \citep{2021ApJ...908..218H}.
However, our obtained value ($\sim$ 0.6) is not much steeper than expected for a linear loss of alignment. This suggests that the grain alignment is being maintained relatively well in this region, perhaps due to the strong asymmetric interaction of the C1 clump with the radiation field of HD 200775. There are also two protostellar cores with Class I (P17 23) and Class II protostars (P17 24) embedded in the high-density clump C1 \citep{pattle2017}. \cite{2001ApJ...561..871H} has found a similar result for CB 54 and DC 253-1.6 with $\alpha$ as 0.64 and 0.55, respectively. Both of these have embedded sources which may be possibly responsible for maintaining the RAT. Alternatively, the presence of either multiple magnetic field components along the line-of-sight or complex magnetic field structure on the scales smaller than the beam may be responsible for the depolarization towards the high-density regions.

\subsection{Magnetic field strength}
We estimated the magnetic field strengths in NGC 7023 using the Davis-Chandrasekhar-Fermi (hereafter DCF) method \citep{davis1951,1953ApJ...118..113C}, which assumes that perturbation in the magnetic field angle from the mean direction of the field arise from small-scale non-thermal motions of the gas i.e. the perturbations in the magnetic fields are Alfv\'{e}nic. 
The conventional methods of estimating magnetic field strength using DCF relation include effects from the turbulent eddies averaged along the line of sight and tend to overestimate the mean plane-of-sky magnetic field by a factor of $\sqrt{N_{turb}}$ \citep{cho2016}, where N$_{turb}$ is the number of independent turbulent cells along the line of sight. Therefore, \cite{cho2016} proposed a modified DCF method for the calculation of the plane-of-sky magnetic field strength: 
\begin{equation}
 B_{pos} = \textit{Q} \sqrt{4 \pi \rho} \frac{\delta v_{c}}{\delta \theta},    
\end{equation}\label{eq:dcf_mod}
where $\sigma_{v}$ is the centroid velocity dispersion of the region in \kms, $\rho$ is the gas density in g cm$^{-3}$, and $\delta\theta$ is the polarization angle dispersion over the selected region. \textit{Q} is the constant of order unity (0.7 $\leq$ \textit{Q} $\leq$ 1) derived through numerical simulations as suggested by \cite{cho2016}. The authors found the ratio of the standard deviation of centroid velocities and the average velocity dispersion of the molecular line data, ($\delta v_{c}$/$\delta$v) is proportional to 1/$\sqrt{N_{turb}}$, where $\delta$v is the average velocity dispersion obtained using molecular line data.

\subsubsection{Structure function}
\begin{figure}
    \centering
    \includegraphics[width=0.5\textwidth]{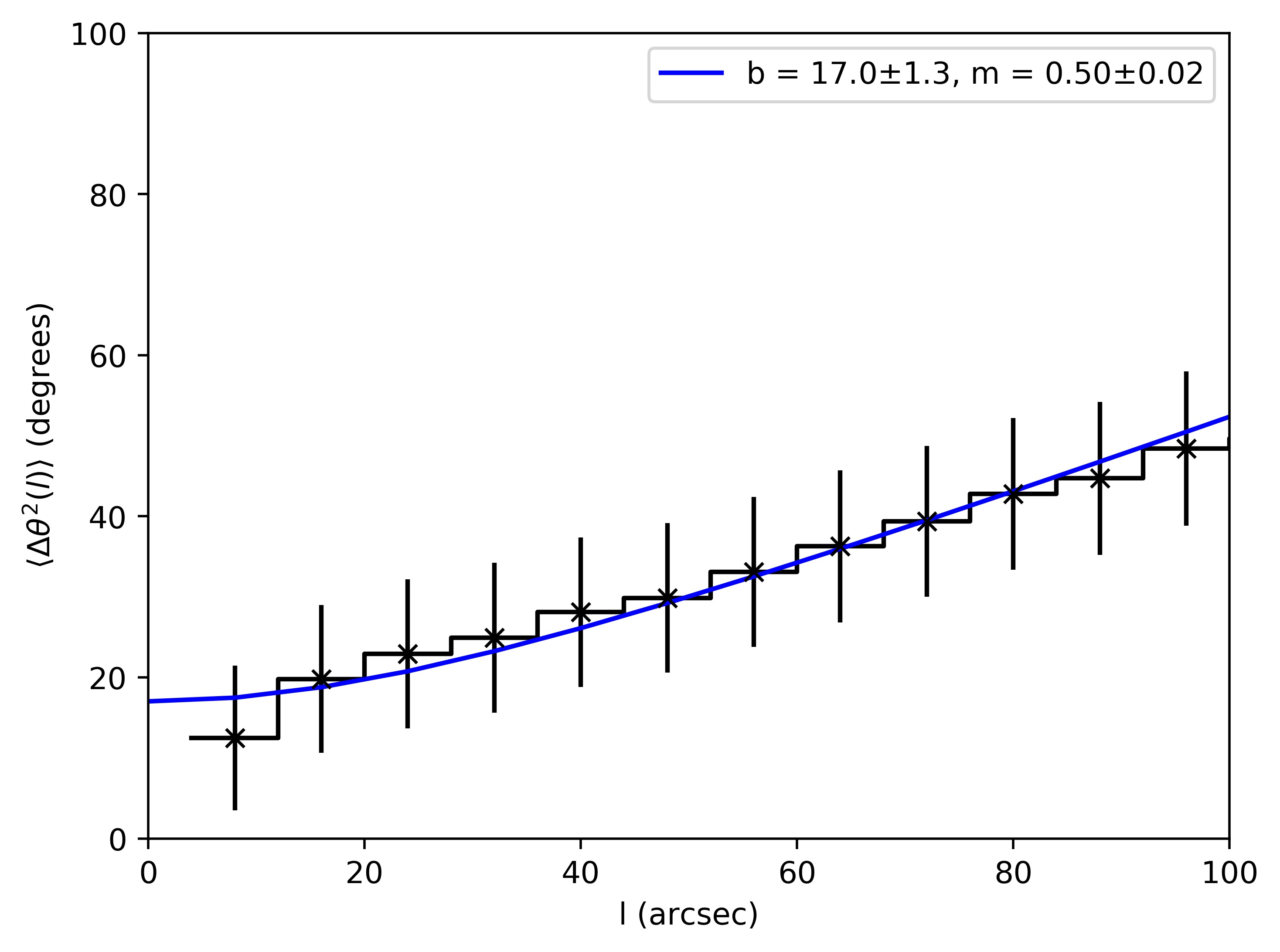}
    \caption{The angular dispersion function for the polarization angles obtained from POL-2 data. The best-fit parameters were found by fitting equation \ref{eq:dispersion} and are shown in the top right legend.}
    \label{fig:sf}
\end{figure}
We estimated magnetic field strengths using the \cite{hildebrand2009} implementation of the DCF method \citep{davis1951, 1953ApJ...118..113C}, in which the ordered and turbulent components of the magnetic field are separated using a structure-function-based approach. 

We calculated the difference in polarization angle between individual pairs of polarization segments, $\Delta \theta (l) \equiv \theta(x) - \theta (x+l)$, where $\theta (x)$ is the polarization angle of a segment at position $x$, and $\theta (x+l)$ is the polarization angle of a segment separated from $x$ by a distance $l$. If the number of pairs is given by $N(l)$, the angular dispersion function is then given by
\begin{equation}
\langle\Delta \theta^2 (l)\rangle^{1/2} \equiv \left[ \frac{1}{N(l)}\sum^{N(l)}_{i=1}\Delta \theta (l)^2 \right] ^{1/2}.
\end{equation}
This can, at small $l$, be fitted with the quadratic function
\begin{equation}
     \langle\Delta \theta ^2 (l)\rangle = b^2 + m^2l^2 +\sigma_M^2(l),
\label{eq:dispersion}
 \end{equation}
where $ml$ arises from the ordered magnetic field (referred to as the large-scale field by \citealt{hildebrand2009}), $b$ is the turbulent dispersion about the mean magnetic field, and $\sigma_M(l)$ is the measurement uncertainty.

Figure~\ref{fig:sf} shows our structure function: $b=17.0\pm 1.3$ degrees ($0.30\pm 0.02$ radians).
The turbulent to large-scale magnetic field strength ratio is given by $b/\sqrt{2-b^2}$. When the turbulent component of the magnetic field is much smaller than the ordered component, i.e. $b\ll1$ rad, $b/\sqrt{2-b^2} \to b/\sqrt{2}$, the DCF relation becomes
\begin{equation} 
B_{pos} = \sqrt{8\pi \rho}\frac{\sigma_{v}}{b},
\label{eq:dcf}
\end{equation}
where $B_{pos}$ is the plane-of-sky magnetic field strength, $\rho$ is gas mass density, and $\sigma_{v}$ is the average non-thermal velocity dispersion of the gas. 

\subsubsection{Gas properties: Velocity dispersion and density}
\begin{figure}[hbt!]
\centering
\includegraphics[height=7cm, width=9cm]{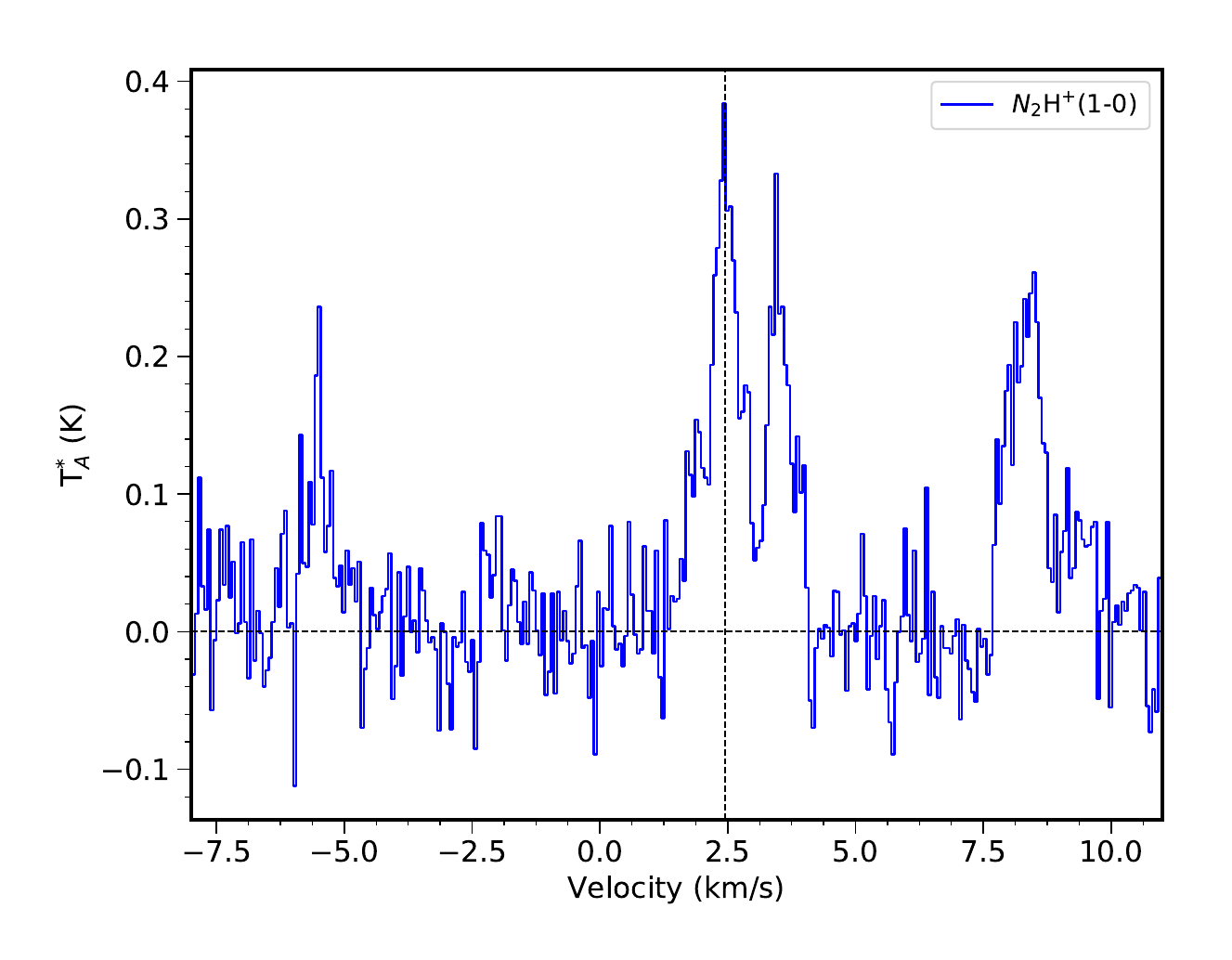}
\caption{Average profile of N$_{2}$H$^{+}$ in blue over the clump. The vertical line shows the systematic velocity of the clump derived using a hyperfine fitting of the \nthp line.}\label{fig:line_nthp}
\end{figure}

Fig. \ref{fig:line_nthp} shows the spectrum of N$_{2}$H$^{+}$ (1$\textendash$0) averaged over the clump in units of antenna temperature (K). We used the \nthp (1$\textendash$0) tracer for estimating the magnetic field strength as it correlates well with the dust continuum emission. We compared the line profiles and velocity dispersion and calculated magnetic field strength using each of the three tracers (Table \ref{tab:dcf}). More details about the choice of the best possible tracer for the field strength analysis are given in Appendix A.

The \nthp line exhibits a hyperfine structure showing seven components, and we obtained the FWHM by fitting Gaussian to the spectra. We fitted seven hyperfine components simultaneously with seven Gaussians using their line parameters given by \cite{1995ApJ...455L..77C}. For our analysis, we considered the velocity dispersion of the central bright component, which is representative of the cloud's systematic velocity. To derive the turbulent line width of the \nthp tracer used in our magnetic field strength calculation, we corrected the measured line widths for thermal broadening with the following relation, with the assumption that the total velocity dispersion is the summation of thermal and non-thermal contributions added in quadrature \citep{1983ApJ...270..105M},
 \begin{equation} 
\sigma_{nt} = \sqrt{(\sigma_{obs})^{2} - (\sigma_{th})^{2}},
\end{equation}
where $\sigma_{th}$ is $\sqrt{kT/\mu m_{obs}}$, thermal velocity dispersion, $\mu$ is the molecular weight of the observed \nthp molecule, T is the gas temperature, and k is the Boltzmann constant.  
We used the dust temperature of the cores as the gas temperature to calculate the thermal line width. \cite{pattle2017} derived the dust temperature for three cores in the range 21.8-23.5 K. Using the average T$_{dust}\sim$ 22.8 K and the mean velocity dispersion of the central bright component as 0.28$\pm$0.07 \kms, we calculated the non-thermal velocity dispersion as 0.26$\pm$0.08 \kms. The beam size of the TRAO data ($\sim$56$^{''}$) is four times higher than that of JCMT, and the cores are not resolved at that angular scale. Therefore, we used the average value of non-thermal velocity dispersion for three cores. 
The gas density, $\rho$ = $\mu_{g}m_{H}$n(H$_{2}$),  where $\mu_{g}$ = 2.8 is the mean molecular weight of the gas, m$_{H}$ is the mass of a hydrogen atom and n(H$_{2}$) is the number density of molecular hydrogen in cm$^{-3}$.  We used the n(H$_{2}$) estimates from \cite{pattle2017} calculated using dust continuum emission, and the values for P17 21, P17 23 and P17 24 are 11.5$\pm$0.7 $\times$ 10$^{4}$ cm$^{-3}$ , 5.3$\pm$0.4 $\times$ 10$^{4}$ cm$^{-3}$ and 8.1$\pm$0.8 $\times$ 10$^{4}$ cm$^{-3}$, respectively. We calculated N$_{turb}$ for \nthp spectra as 3.4 using the \cite{cho2016} method described above. Using the n(H$_{2}$), $\sigma_{nt}$ and b (shown in Table \ref{tab:dcf_nthp}), we thus calculated the magnetic field strength for three cores using equation \ref{eq:dcf} and further multiplied it by 1/$\sqrt{N_{turb}}$.

We calculated the uncertainties in the B-field strength using the propagation of errors in the other quantities:
\begin{equation}
   \delta B_{pos} = B_{pos}\sqrt{\left(\frac{1}{2} \frac{\delta n_{H_{2}}} {n_{H_{2}}}\right)^{2} + \left(\frac{\delta \sigma_{nt}} {\sigma_{nt}}\right)^{2} + \left(\frac{\delta b}{b}\right)^{2}},
\end{equation}
where $\delta$n(H$_{2}$), $\delta\sigma_{nt}$, and $\delta$b are the uncertainties in the number density, non-thermal velocity dispersion, and the polarization angle dispersion, respectively. 
Finally, we estimated the magnetic field strengths using the modified DCF method as 179$\pm$50 $\mu$G, 121$\pm$34 $\mu$G, and 150$\pm$42 $\mu$G for P17 21, P17 23, and P17 24, respectively.

\subsection{Mass-to-flux ratio and magnetic criticality of the clump}
The stability of the clump can be assessed using the mass-to-flux ratio, $\lambda$. $\lambda<$1 implies that the clump is magnetically supported, while if $\lambda>$1, it implies that the clump is unstable to gravitational collapse and is not supported well by the magnetic field. We estimated the mass-to-flux ratio \citep{2004Ap&SS.292..225C}: 
\begin{equation}
    \lambda = 7.6\times10^{-21}\times N_{{H}_{2}} (cm^{-2})/B_{pos} (\mu G),
\end{equation}
 where N$_{H_{2}}$ is the column density, and B$_{pos}$ is the plane-of-the-sky magnetic field strength.  The dust column density values, N$_{H_{2}}$ estimated using SCUBA emission are (13.7$\pm$0.9) $\times 10^{21}$ cm$^{-2}$, (7.3$\pm$0.6) $\times 10^{21}$ cm$^{-2}$, and  (11.2$\pm$1.0) $\times 10^{21}$ cm$^{-2}$ for P17 21, P17 23, and P17 24, respectively \citep{pattle2017}. Using these N$_{H_{2}}$ and our B$_{pos}$ estimates from \nthp lines, we estimated $\lambda$ as $0.6\pm0.2$, $0.5\pm0.1$, and $0.6\pm0.2$ for cores P17 21, P17 23 and P17 24, respectively. These values suggest that the starless P17 21 and protostellar cores, P17 23 and P17 24, are magnetically subcritical, implying that they are supported against gravitational collapse by their magnetic fields. However, it should be noted that two of the cores, P17 23 and P17 24, host embedded protostars and, therefore, are not stable. It is important to note that the values of $\lambda$ qualitatively suggest the dynamical importance of magnetic fields in their evolution without precisely measuring the stability against collapse and that our values of B-field strength might also be overestimated \citep{2023ASPC..534..193P}. 

\begin{table*}
    \centering
    \caption{DCF results and energetics determined using the N$_2$H$^+$ (1$\textendash$0) line}
    \begin{tabular}{ccc  ccc ccc ccc}
    \hline
        Core & $n({\rm H}_{2})$ & $b$ & $\sigma_{v,\textsc{nt}}$ & $N_{turb}$ & $B$ 
        & $\lambda$& $E_{P}\dagger$ & $E_{G}$ & $E_{B}$ & $E_{outflow}$ & Type\\
         & ($\times 10^4$ cm$^{-3}$) & (degrees) & (km\,s$^{-1}$) & & ($\mu$G) & & $\times10^{39}$ & $\times10^{42}$ & $\times10^{42}$  & $\times10^{42}$  & \\
    \hline
        21 & $11.5\pm 0.7$ & $17.0\pm 1.7$ & $0.26\pm0.08$ & 3.4 & $179\pm50$ &$0.6\pm0.2$ & 1.5 & 0.4 & 3.8 & 3.9 & Starless\\
     23 & $5.3 \pm 0.4$ & $17.0\pm 1.7$ & $0.26\pm0.08$ & 3.4 & $121\pm34$ & $0.5\pm0.1$ &2.3 & 1.0 & 2.8 & 5.9 & Protostellar\\ 
     24 & $8.1\pm 0.8$ & $17.0\pm 1.7$ & $0.26\pm0.08$ & 3.4 & $150\pm 42$ &$0.6\pm0.2$ &2.3 & 0.7 & 4.3 & 5.9 & Protostellar \\
    \hline
    \end{tabular}
    $\dagger${Energy values are in the units of ergs}
    \label{tab:dcf_nthp}
\end{table*}

\section{Discussion}
In this paper, we present magnetic field estimates in a reflection nebula triggered by an intermediate-mass star that acts as a bridge between low- and high-mass star-forming regions \citep{fuente1998}. The region of our study, NGC 7023, is a good example of triggered star formation having a pre-main sequence Herbig Ae Be star, HD 200775, as the central source which has triggered the formation of many dense condensations, some of which are forming young stellar objects (YSOs) \citep{2009ApJS..185..198K}. There is a biconical cavity in the east-west direction, which is likely to have been carved out by a bipolar outflow in an earlier evolutionary stage and is currently chemically inactive \citep{1998A&A...339..575F}. The age of the central star HD 200775 is 1-3 Myr \citep{2018A&A...620A.128V}, which is much longer than the lifetime of the Class I and Class II YSOs clustered around the hub with ages of around 0.1 Myr. The nebula can be considered a sheet of dense molecular gas in which a massive star is born and subsequently disperses the surrounding gas through radiative and stellar feedback. The pre-main sequence stars with luminosities less than massive stars \citep{1993ApJ...418..414P} cannot create extended \hii~ regions; however, in several cases, extended atomic gas regions can be found associated with them \citep{fuente1998}. The pre-main sequence star, HD 200775, with bolometric luminosity, L$_{bol}\sim$ 8000 L$_{\odot}$ \citep{1997A&A...324L..33V} and mass 10 M$_{\odot}$, has created three PDRs, or bright rims, located at the edges of the molecular cloud surrounding it and also the extended atomic regions \citep{1998A&A...334..253F}. We focus on the clump located near PDR 1 (see left panel of \ref{fig:stokes_I}). The polarization observations using JCMT POL-2, with a linear scale of 0.024 pc, show an aligned magnetic field structure along the high-density structure (black skeleton in Figure 3). The magnetic field changes its orientation from being perpendicular at large-scale (measured by \textit{\textit{Planck}}/optical) to parallel to the high-density structure at small scales (0.5 pc to 0.03 pc). This is opposite to what we expect in gravitationally dominated and magnetically supercritical regions, with the magnetic fields observed to be perpendicular to the clump major axis. This hints towards the role of an external factor in regulating the dynamics of this high-density clump. Previous studies investigating the magnetic fields near PDRs in nearby and distant star-forming regions have identified similar reordered field structures \citep{WardThompson2017, Pattle2018,Kwon2018,Hwang2023}. 

\subsection{Energetic analysis and possible effect of the stellar feedback on the polarization}
 To investigate the physical factors involved further, we estimated the energy budget of the cloud.

We calculated the magnetic energy ($E_{B}$) of the clump and compared it with other energy terms: gravitational ($E_{G}$), photon ($E_{P}$), and outflow ($E_{O}$) energies integrated over the whole region around it. 

\subsubsection{Magnetic energy}
We calculated the magnetic energy of the three cores as 
\begin{equation}
    E_{B} = \frac{4}{3} \pi R^{3} B^{2}_{pos}/8\pi, 
\end{equation}
where B$_{pos}$ is the plane-of-sky magnetic field strength calculated from DCF analysis. 
The calculated magnetic energies are 3.8 $\times$ 10$^{42}$ ergs, 2.8 $\times$ 10$^{42}$ ergs and 4.3 $\times$ 10$^{42}$ ergs for P17 21, 23 and 24, respectively.
 
\subsubsection{Radiation field energy}
We first estimated the total rate at which ionizing photons are emitted as N$_{LyC}$ = 4$\pi R^2 N^{0}_{LyC}$, where the stellar radius is assumed to be 4.6 R$_{\odot}$ and the number of Lyman continuum photons emitted per unit surface area to be 10$^{20.4}$ cm$^{-2}$ s$^{-1}$. The incident radiation field depends on the spectral type and the geometry of the ionizing star. The spectral type of HD 200775 is similar to HD 147889, and so we adopted the stellar properties - radius and Lyman continuum photons from \cite{2015MNRAS.450.1094P}.

We estimated the mean radiation energy from the ionizing photons emitted by the ionizing central star of spectral type B2V and penetrating the high-density cloud using the relation, 
\begin{align}
E_{P} = \frac{4R^{2}k_{B}T_{H\textsc{ii}}}{D_{H\textsc{ii}}} (\frac{3\pi N_{LyC} R}{\alpha})^{1/2},
\end{align}
where T$_{H\textsc{ii}}$ is the canonical temperature of the gas in an H\textsc{ii} region, and D$_{H\textsc{ii}}$ is the projected distance of the core from the central star.
The projected distance of the clump from HD 200775 is 0.05 pc, assuming a distance of 340 pc \citep{2020MNRAS.494.5851S}. The calculated photon energies are 1.5 $\times$ 10$^{39}$ ergs, 2.3 $\times$ 10$^{39}$ ergs and 2.3 $\times$ 10$^{39}$ ergs for P17 21, 23 and 24, respectively.

\subsubsection{Gravitational Energy}
We estimated the gravitational potential energy of the cores by assuming a spherically symmetrical density distribution with radius R using the relation,
\begin{equation}
    E_{G} = -\frac{3}{5} \frac{GM^{2}}{R}.
\end{equation} We used the projected deconvolved radii of the cores as 0.029, 0.034 and 0.034 pc from \cite{pattle2017}. Therefore, the calculated gravitational energies are 0.4 $\times$ 10$^{42}$ ergs, 1.0 $\times$ 10$^{42}$ ergs and 0.7 $\times$ 10$^{42}$ ergs for P17 21, 23 and 24, respectively.
\begin{figure}[ht!]
    \centering
\includegraphics[width=9cm, height=8cm]{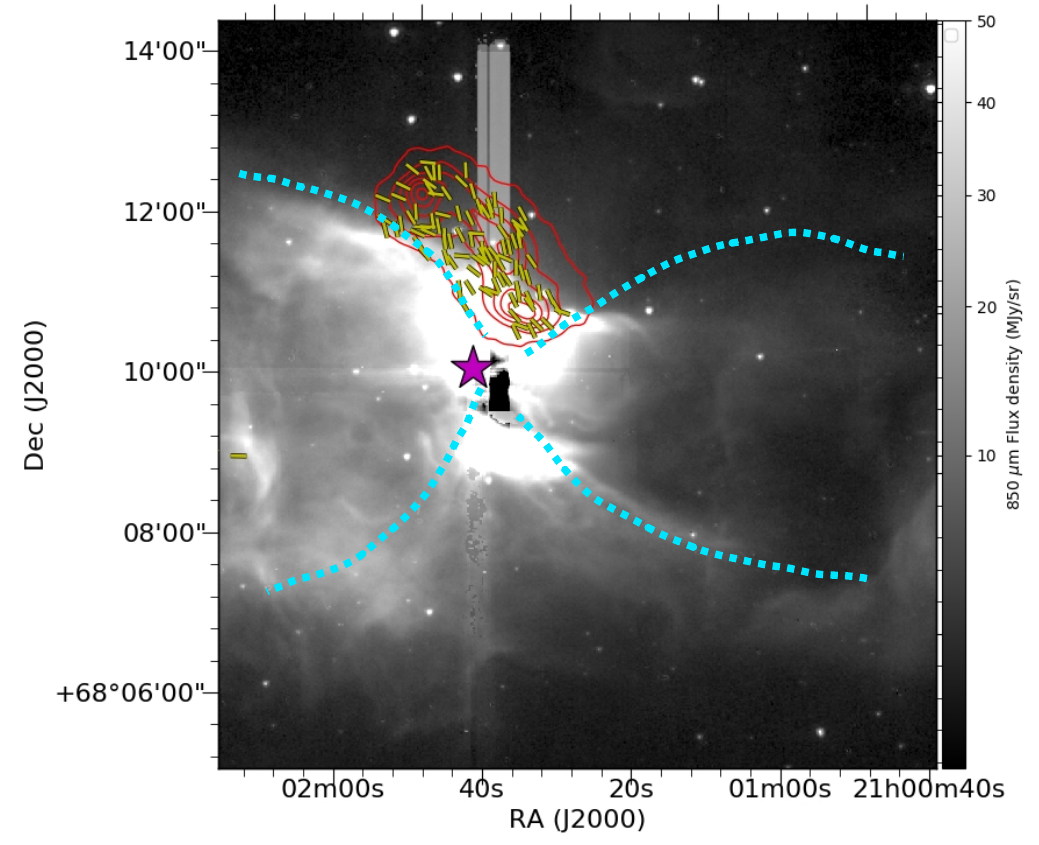}
    \caption{\textit{Spitzer} 8 $\mu$m emission map of NGC 7023 taken from \textit{c2d} catalog \citep{2003PASP..115..965E}. The red contours and yellow vectors show the SCUBA-2 emission contours and the magnetic field orientation derived from our POL-2 measurements, respectively. The dotted cyan curves mark the outer periphery of the biconical structure of the nebula, and the magenta star shows the central source HD 200775. The extent of the emission region around HD 200775 corresponds to the yellow box marked in Fig. \ref{fig:ngc_rgb}.}
    \label{fig:spitzer_ngc}
\end{figure}

\begin{figure*}[ht!]
    \centering
\includegraphics[width=\textwidth,height=9cm]{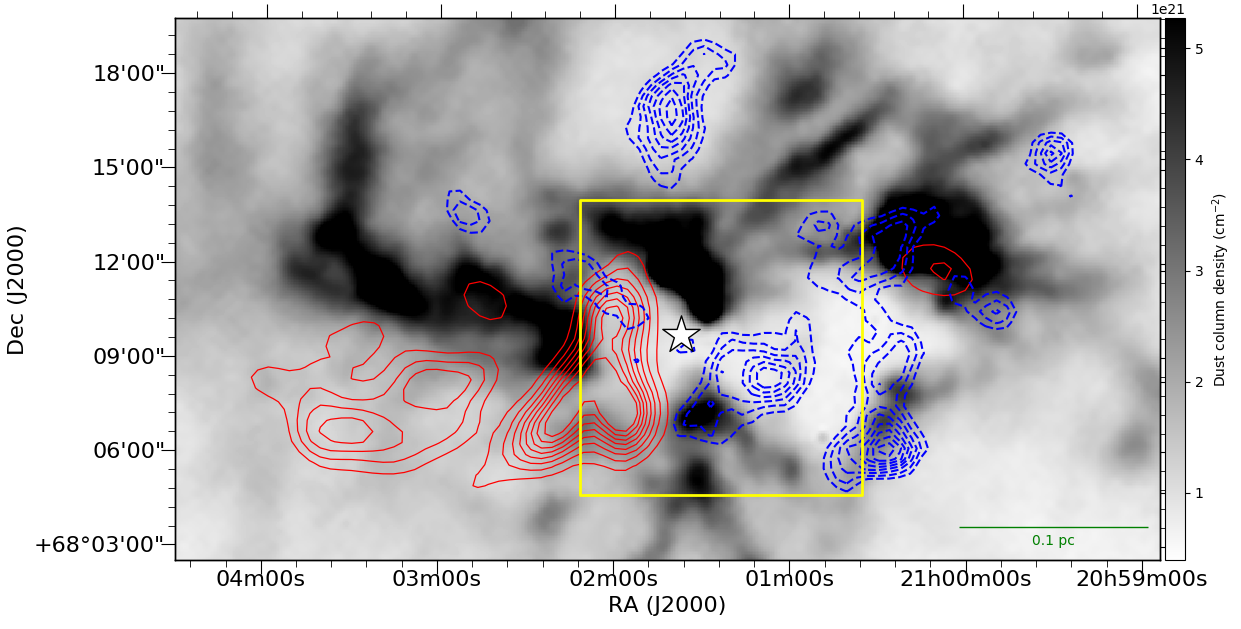}
    \caption{$\textit{Herschel}$ dust column density map of the region with red- and blue-shifted high velocity \co gas around the star. The red contours show the integrated \co emission in the velocity range 5 to 7.5 \kms and blue dotted contours in the velocity interval -2.5 to 0 \kms.
    The contour levels for red contours range from 1.25 to 7 K \kms in steps of 5$\sigma$ with $\sigma$ as 0.15 K \kms, and for blue-shifted gas, the levels range from 1.3 to 6 K \kms in steps of 3$\sigma$ ($\sigma\sim$0.18 K \kms). The yellow box marks the extent of the \textit{Spitzer} data shown in Fig. \ref{fig:spitzer_ngc}. The systematic velocity of the cloud is $\sim$2.7 \kms.}
\end{figure*}\label{fig:ngc_rgb}

\begin{figure*}
    \centering
\includegraphics[width=14cm, height=10cm]{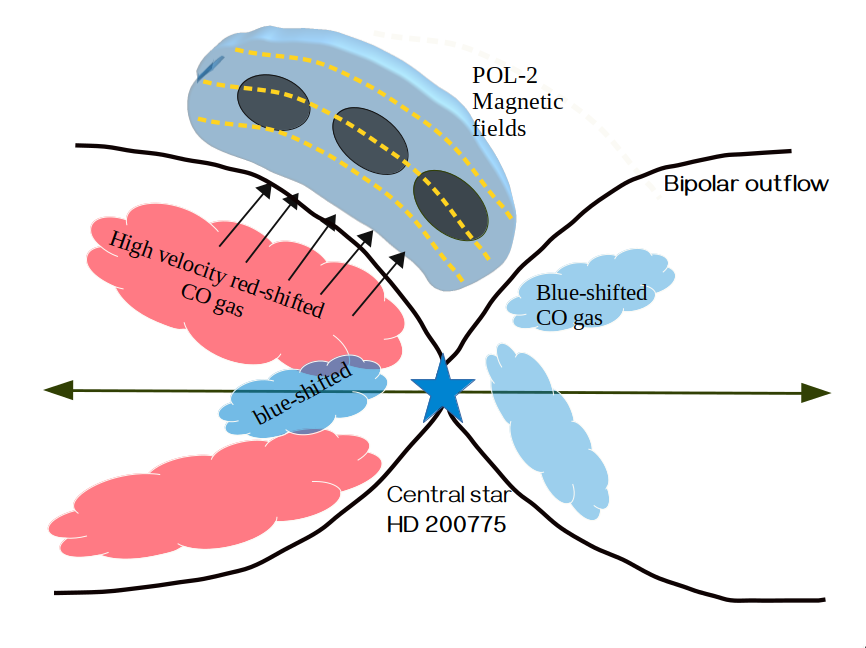}
    \caption{Schematic view of the clump C1 in the NGC 7023 region. Black solid lines show the bipolar outflow, and the double arrow shows the direction of the outflow axis. The central source HD 200775 is marked by a blue star. The black ellipses show three cores in the blue-coloured clump. The dashed lines show the derived magnetic field direction in the clump using POL-2. Black arrows show the direction of the compression across the outflow walls. The red and blue structures show the red- and blue-shifted gas traced in outflow wings.}
\end{figure*}  \label{fig:scheme_ngc}

\subsubsection{Kinetic energy of outflowing material}
Fig. \ref{fig:spitzer_ngc} shows the \textit{Spitzer} 8 $\mu$m emission map obtained from the \textit{c2d Spitzer} Legacy project data release\footnote{\url{http://ssc.spitzer.caltech.edu/legacy/c2dhistory.html}}. The magnetic field orientation towards the clump can be seen at the periphery of the reflection nebula (dotted lines in cyan colour). The clump's location and the nebula's geometry suggest that the outflow from the central source has significantly reshaped the clump. The systematic velocity of the nebula using \coo (1-0) averaged over the whole nebula is $\sim$2.7 \kms (Sharma et al. in preparation).  Fig. \ref{fig:ngc_rgb} shows the distribution of high-velocity gas in the red- and blue-shifted wings, which is overplotted on the \textit{Herschel} dust column density map of NGC 7023. To understand the effect of outflow on the clump energetics, we estimated the outflow kinetic energy using \co (1$\textendash$0) emission in the high-velocity ranges of 4-5 \kms and 5-7.5 \kms. Then, we considered its possible contribution to the energy balance of each of the three cores. The H$_{2}$ mass and the kinetic energy, $\frac{1}{2}$M$_{outflow}$v$^{2}$ are calculated to be $\sim$ 37 M$_{\odot}$ and 1.6$\times$10$^{45}$ ergs in the velocity range of 4-5 \kms, and 29 M$_{\odot}$ and 6.2$\times$10$^{45}$ ergs in the range of 5-7.5 \kms. \citet{fuente1998} modelled the outflow as a pair of cones, each of diameter 0.9\,pc and slant height 1\,pc. Again, adjusting for the difference in assumed distance, this is equivalent to a surface area of 1.28\,pc$^{2}$. We assume that the energy of the outflow is distributed evenly over this surface area, although we note that this likely results in an underestimate of the true energy incident on our cores, as they are very close to the exciting star. \cite{pattle2017} derived FWHMs for the three cores of 0.029, 0.035 and 0.035 pc. The areas projected by these three cores onto the cone are taken to be $\pi\times({\rm FWHM}/2)^{2}$, 0.052\%, 0.075\% and 0.075\% of the total surface area, respectively. Thus, we estimate outflow kinetic energies incident on our cores of $\gtrsim 3.9\times 10^{42}$\, erg for core P17 21, and $\gtrsim 5.9\times 10^{42}$\,erg for cores P17 23 and P17 24, comparable to the gravitational and magnetic energies that we estimate for the cores. 

 Fig. \ref{fig:scheme_ngc} shows a schematic of NGC 7023, summarizing the suggested scenario with the B-field structure towards C1 and the outflow feedback through high-velocity gas shown. The dashed yellow lines show the B-field geometry of the clump (with marked cores in black), which is perpendicular to the direction of the ionization radiation. \cite{Fuente1996} showed the high-density filaments/clumps around the bipolar outflow could have been created by its interaction with the surrounding cloud. Using the hydrodynamic simulations of magnetized clouds under the radiative feedback,  \cite{2011MNRAS.414.1747A} showed that the \hii~region sweeps the material while expanding due to the bow shock front. This expansion of \hii~region compresses the high-density gas, dragging the magnetic field lines due to flux freezing conditions and aligning the magnetic field along the high-density clump. \cite{2011MNRAS.412.2079M} also further investigated the structural transformation of globules with different magnetic field strengths where the initially perpendicular B-field will get altered and enhanced. Expansion of \hii~regions and the compression of the molecular gas in the surrounding cloud thus helps increase the density of the clump, thereby inducing the fragmentation of the cloud.
We suggest that the reordered and enhanced B-field strength of C1 in NGC 7023 is a consequence of the outflow feedback (see Figure \ref{fig:spitzer_ngc} and \ref{fig:ngc_rgb}), where the initial weak B-fields get compressed through the PDR expansion, and the B-field strength is thus amplified ($\sim$150-179 $\mu$G) in the clump C1.

However, it is important to note that the comparable energies of the cores suggest the system is near equilibrium, whereas there is ongoing star formation in two cores, P17 23 and P17 24. We suggest that the magnetic pressure in C1 would keep increasing in a contrary response to the expanding inward momentum of the PDR layers until the energies are comparable and the system attains equilibrium. This prevents further dissociation due to stellar feedback and maintains the evolution of the clump in a strong field regime.
In addition, the magnetic field strengths are also likely to be systematically overestimated \citep{2023ASPC..534..193P}, which will affect the estimation of the magnetic energies of the cores. While we have used the \cite{cho2016} method to correct the magnetic field strength, \cite{2021ApJ...919...79L} suggested that this method can only account for the effect of line-of-sight integration for the scales higher than 0.1 pc. The role of the B-field might, therefore, be dominant only at the clump scales but subdominant in affecting 
the cores’ stability against gravitational collapse. Future high-resolution polarization observations towards the cores will be helpful in gaining insights into the role of magnetic fields in cores' stability.  
 
\section{Conclusions}
We have presented the polarized dust emission observations of NGC 7023, a reflection nebula also named \textit{Iris} nebula, using the POL-2 polarimeter on the JCMT at 850 $\mu$m. We found reordered magnetic fields parallel to the high-density structure. We modelled the relationship between polarization fraction and the dust emission intensity. We found the power index to be 0.65, which implies a decrease in the dust grain alignment efficiency with increasing A$_{V}$.
We also compared our POL-2 results with previous starlight and \textit{\textit{Planck}} dust polarization and found that POL-2 shows a completely different distribution of polarization vectors, suggesting a different physical mechanism influencing the magnetic field on smaller scales within the nebula. The magnetic field strengths are estimated using \nthp lines observed with the TRAO telescope, and the calculated values are 179$\pm$50 $\mu$G, 121$\pm$34 $\mu$G, and 150$\pm$42 $\mu$G for P17 21, P17 23, and P17 24, respectively. Both the protostellar cores, P17 23 and P17 24, and the starless core, P17 21, are magnetically subcritical, implying the dominating role of magnetic fields, but the ongoing star formation in the cores suggests our magnetic field strengths might be overestimated.
The clump direction and the walls of the outflow cavity on the northwest side of the star are aligned, and the high-density gas shows the signature of interaction with the high-velocity gas. We conclude that the B-fields in the clump have been compressed and reordered by the outflow feedback.

\medskip
\section*{Acknowledgements}
 This work is supported by the National Natural Science Foundation of China (NSFC)(Grant No. 11988101) and by the Alliance of International Science Organizations, Grant No. ANSO-VF-2021-01. We are grateful to the anonymous referee for their valuable comments that have improved the presentation of this paper. D.L. is a new cornerstone investigator. K.P. is a Royal Society University Research Fellow, supported by grant number URF\textbackslash R1\textbackslash 211322. C.W.L. acknowledges support from the Basic Science Research Program through the NRF funded by the Ministry of Education, Science and Technology (NRF- 2019R1A2C1010851) and from the Korea Astronomy and Space Science Institute grant funded by the Korea government (MSIT; project No. 2024-1-841-00). The James Clerk Maxwell Telescope is operated by the East Asian Observatory on behalf of the National Astronomical Observatory of Japan; the Academia Sinica Institute of Astronomy and Astrophysics; the Korea Astronomy and Space Science Institute; the National Astronomical Research Institute of Thailand; the Center for Astronomical Mega-Science (as well as the National Key R\&D Program of China with No. 2017YFA0402700). Additional funding support is provided by the Science and Technology Facilities Council of the United Kingdom and participating universities and organizations in the United Kingdom, Canada and Ireland. Additional funds for the construction of SCUBA-2 were provided by the Canada Foundation for Innovation. The authors wish to recognize and acknowledge the very significant cultural role and reverence that the summit of Maunakea has always had within the indigenous Hawaiian community.  We are most fortunate to have the opportunity to conduct observations from this mountain. This research has made use of the NASA/IPAC Infrared Science Archive, which is funded by the National Aeronautics and Space Administration and operated by the California Institute of Technology.

\appendix
\section{Calculation of DCF strengths for three tracers}
\begin{figure}[hbt!]
\centering
\includegraphics[height=8cm, width=10cm]{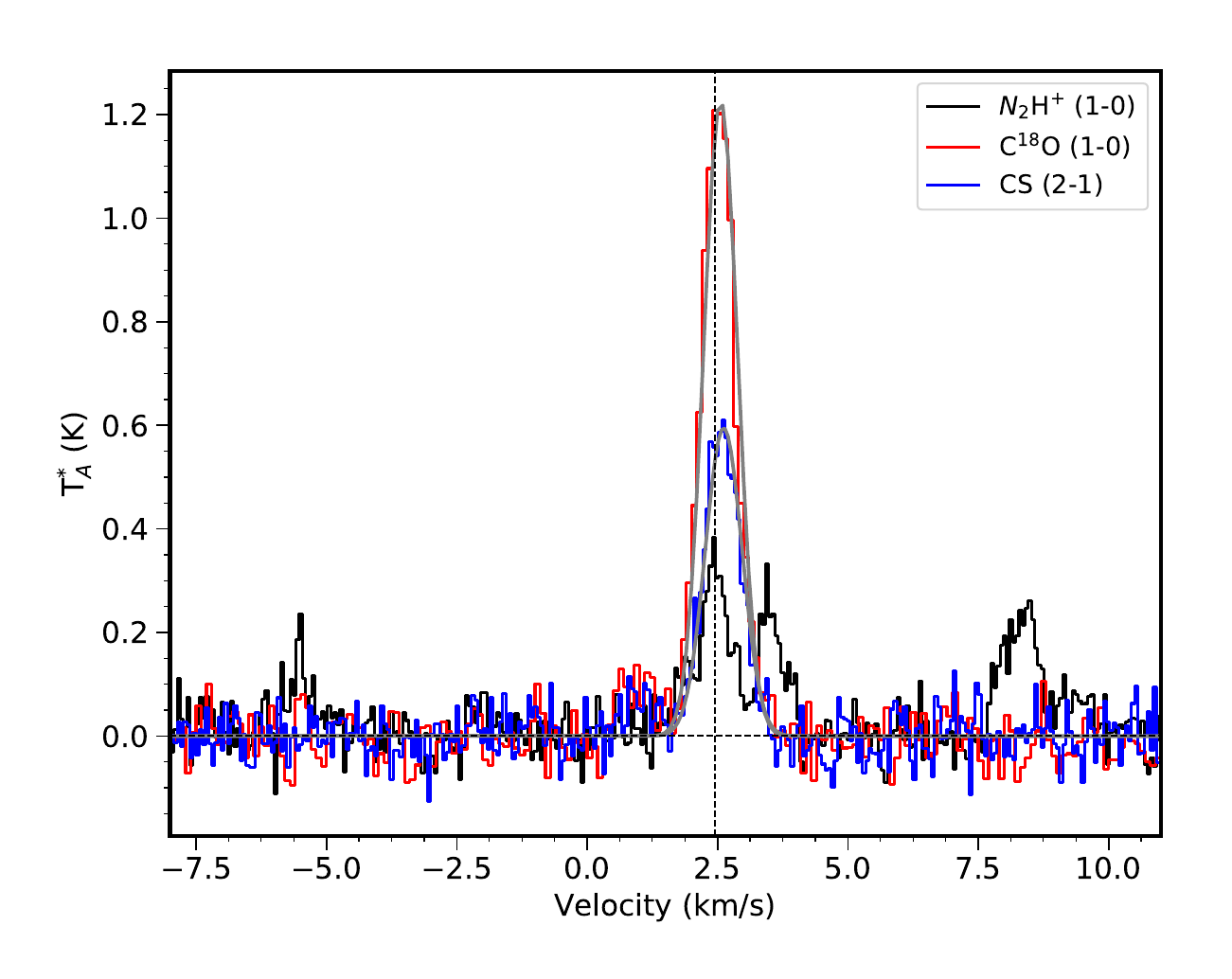}
\caption{Average profiles of N$_{2}$H$^{+}$ in black, \coo in red, and CS (2$\textendash$1) in blue colour towards the clump. The grey line shows the Gaussian fitted. The vertical line shows the systematic velocity of the clump derived using a hyperfine fitting of the \nthp line.}\label{fig:av_n2h_cs_c18o}
\end{figure}

Fig. \ref{fig:av_n2h_cs_c18o} shows the comparison of the spectra of the three tracers, C$^{18}$O (1$\textendash$0), CS (2$\textendash$1), and N$_{2}$H$^{+}$ (1$\textendash$0), averaged over the clump morphology in units of antenna temperature (K). We estimated the velocity dispersion of three different molecular lines using the Gaussian fitting of C$^{18}$O (1$\textendash$0) and CS (2$\textendash$1), and the hyperfine fitting of N$_{2}$H$^{+}$ lines. The mean FWHM of the C$^{18}$O and CS (2$\textendash$1) line is 1.2 and 1.8 km s$^{-1}$, respectively. The C$^{18}$O line traces the high-density gas toward the cloud, while the CS (2$\textendash$1) and \nthp~lines trace the deepest regions of the core, as the critical density is higher for \nthp (n $\sim$ 10$^{6}$ cm$^{-3}$).

\begin{table*}[ht!]
    \centering
    \caption{Comparison of magnetic field strength for three line tracers}
    \begin{tabular}{ccc  ccc ccc ccc}
    \hline
     & & & \multicolumn{3}{c}{N$_2$H$^+$} & \multicolumn{3}{c}{CS} & \multicolumn{3}{c}{C$^{18}$O} \\ \cline{4-6} \cline{7-9} \cline{10-12}
        Core & $n({\rm H}_{2})$ & $b$ & $\sigma_{v,\textsc{nt}}$ & $N_{turb}$ & $B$ & $\sigma_{v,\textsc{nt}}$ & $N_{turb}$ & $B$ & $\sigma_{v,\textsc{nt}}$ & $N_{turb}$ & $B$ \\
         & ($\times 10^4$ cm$^{-3}$) & (degrees) & km\,s$^{-1}$ & & ($\mu$G) & (km\,s$^{-1}$) & & ($\mu$G) & (km\,s$^{-1}$) & & ($\mu$G) \\
    \hline
         21 & $11.5\pm 0.7$ & $17.0\pm 1.7$ & $0.26\pm0.08$ & 3.4 & $179\pm50$  & $0.33\pm0.03$ & 12.1 & $118\pm 15$  & $0.31\pm 0.03$ & 15.1 & $100\pm 14$ \\
         23 & $5.3 \pm 0.4$ & $17.0\pm 1.7$ & $0.26\pm0.08$ & 3.4 & $121\pm34$  & $0.33\pm0.03$ & 12.1  & $80\pm 11$  & $0.31\pm 0.03$ & 15.1 & $68\pm 10$ \\
         24 & $8.1\pm 0.8$ & $17.0\pm 1.7$ & $0.26\pm0.08$ & 3.4 & $150\pm 42$  & $0.33\pm0.03$ & 12.1 & $99\pm 13$  & $0.31\pm 0.03$ & 15.1 & $84\pm 12$ \\
    \hline
    \end{tabular}
    \label{tab:dcf}
\end{table*}

Table \ref{tab:dcf_nthp} shows the calculated magnetic field strengths, estimated using three tracers. The ambiguity of the choice of tracers makes the field strength more uncertain, leading to an underestimation of the stability of the cores against gravitational collapse. Therefore, one needs to be cautious while choosing the line tracer for the strength calculations and stability analysis. 

\section{Correlation of dust emission with \nthp (1$\textendash$0), \coo (1$\textendash$0) and CS (2$\textendash$1) and  emission maps}
\begin{figure}[hbt!]
\centering
\includegraphics[height=7cm, width=7.5cm]{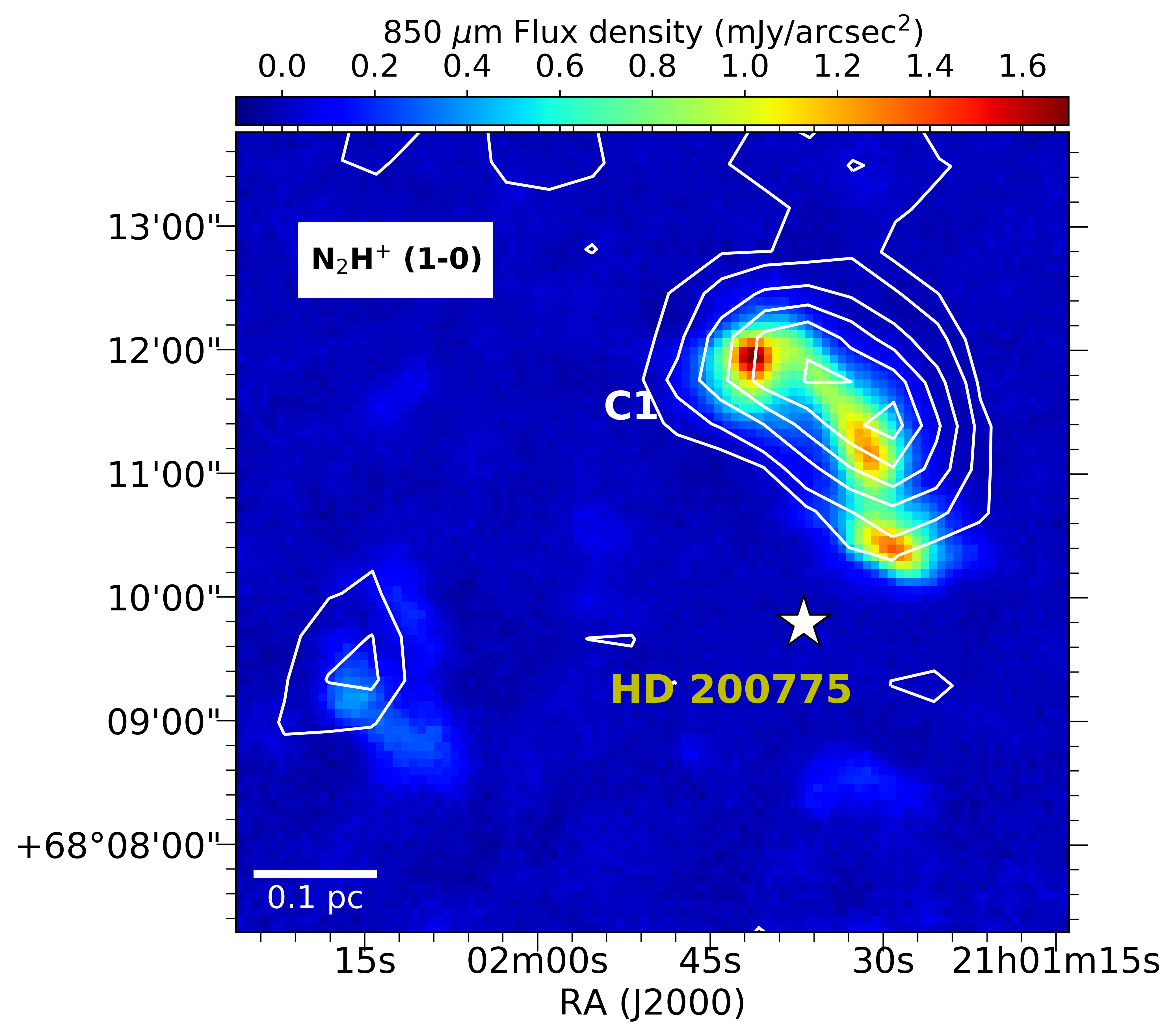}
\includegraphics[height=7cm, width=7.5cm]{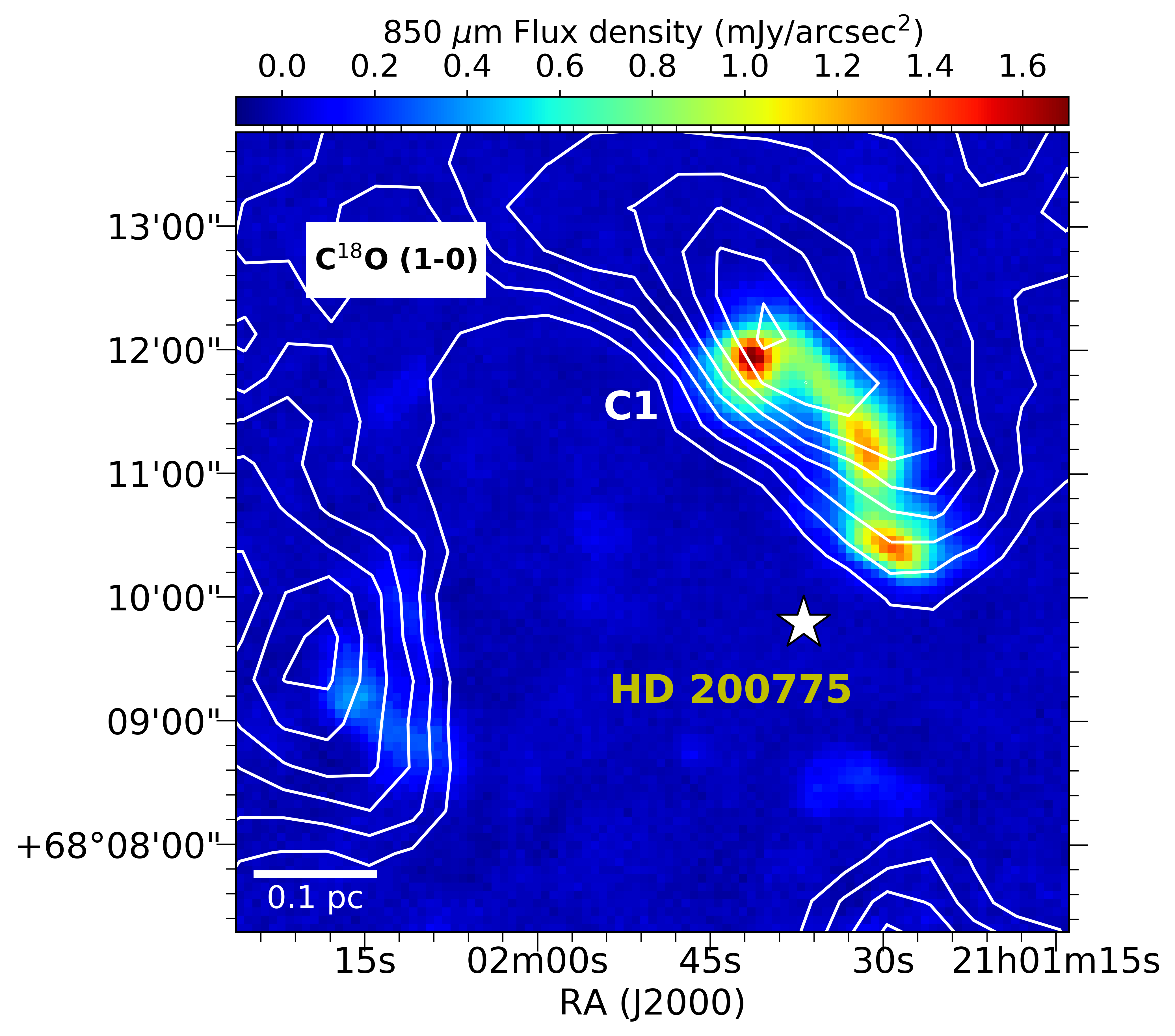}
\includegraphics[height=7cm, width=7.8cm]{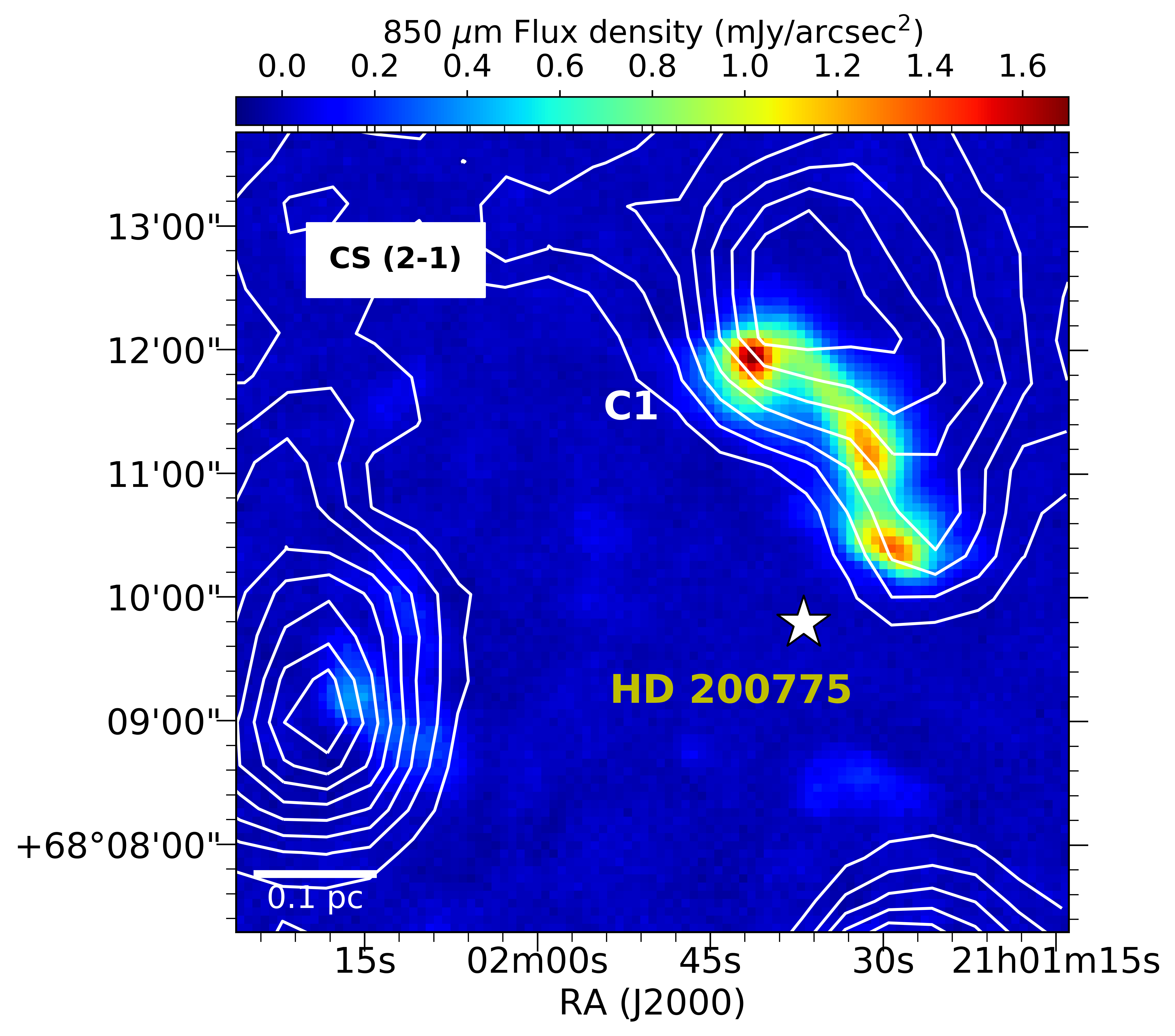}
\caption{SCUBA-2 dust continuum emission map of the observed region with the overplotted integrated molecular line emission contours for (a) \nthp(1$\textendash$0), with the contour levels ranging from 0.47 to 1.3 K \kms in steps of $\sim$0.1 K \kms; (b) \coo (1$\textendash$0) line and the levels range from 0.3 to 1.3 K \kms in steps of 0.14 K \kms; (c) CS(2$\textendash$1) with the contour levels ranging from 0.2-0.8 K \kms in steps of 0.08 K \kms.}\label{fig:mom_0}
\end{figure}

To choose the best possible tracer for estimating magnetic field strength, we compared the integrated line emission of three tracers with the dust continuum emission obtained with SCUBA-2. Fig. \ref{fig:mom_0} shows the moment maps of NGC 7023 using the \nthp(1$\textendash$0), \coo (1$\textendash$0), and CS (2$\textendash$1). The molecular line contours have been overplotted on the dust continuum emission maps of SCUBA-2. We find that the gas and dust emissions are more morphologically correlated for \nthp in panel (c). Therefore, the dust emission and gas traced by an optically thin tracer, \nthp (1$\textendash$0), effectively represent the same layer of cloud. We, therefore, use the \nthp(1$\textendash$0) line to estimate the magnetic field strength in our calculations.

\bibliography{NGC7023}{}
\bibliographystyle{aasjournal}

\end{document}